\documentclass[11pt]{article}

\usepackage[letterpaper,margin=1in,headheight=14pt,headsep=20pt,footskip=30pt]{geometry}
\usepackage[T1]{fontenc}
\usepackage{mathpazo}              % Palatino math
\usepackage{tgpagella}             % Palatino text (TeX Gyre)
\usepackage{tgheros}               % sans for headings and running heads

\usepackage{amsmath,amsfonts,bm}

\def\eqref#1{equation~\ref{#1}}
\def\1{\bm{1}}

\DeclareMathAlphabet{\mathsfit}{\encodingdefault}{\sfdefault}{m}{sl}
\SetMathAlphabet{\mathsfit}{bold}{\encodingdefault}{\sfdefault}{bx}{n}

\usepackage[dvipsnames]{xcolor}
\usepackage{booktabs}
\usepackage{float}
\usepackage{wrapfig}
\usepackage{graphicx}
\usepackage{microtype}
\usepackage{tabularx}
\usepackage{pifont}
\usepackage{natbib}
\usepackage{url}
\usepackage[most]{tcolorbox}
\usepackage{ragged2e}
\tcbuselibrary{raster,skins}
\usepackage{enumitem}
\usepackage{fancyhdr}
\usepackage{titlesec}
\usepackage[font=small,labelsep=period]{caption}
\usepackage{hyperref}

\def \ie {\emph{i.e.}}
\def \eg {\emph{e.g.}}
\def \etal {\emph{et al.}}

\floatstyle{ruled}
\newfloat{algorithm}{tbp}{loa}
\floatname{algorithm}{Algorithm}
\floatstyle{plain}

\renewenvironment{wrapfigure}[3][]{%
  \begin{figure}[!hb]\keeppositivevspace%
  \centering\begin{minipage}{0.62\textwidth}}{\end{minipage}\end{figure}}
\renewenvironment{wraptable}[3][]{%
  \begin{table}[!htb]\keeppositivevspace%
  \centering\begin{minipage}{\dimexpr #3*11/10\relax}\centering}{\end{minipage}\end{table}}

\definecolor{boxheader}{RGB}{40, 50, 65}
\definecolor{arxaccent}{RGB}{31, 78, 140}
\definecolor{arxmuted}{RGB}{104, 112, 124}
\definecolor{arxpanel}{RGB}{243, 245, 248}
\definecolor{arxrule}{RGB}{206, 211, 219}
\usepackage[final]{labreview}
\usepackage{tikz}
\usetikzlibrary{arrows.meta,fit,backgrounds,positioning,calc}
\newcommand{\Constitution}{\mathcal{C}}
\newcommand{\Policy}{\mathcal{P}}

\newcommand{\ConfigSpace}{\Psi}

\newtcolorbox{qualemail}[3][]{
  enhanced,
  colback=white,
  colframe=#3,
  colbacktitle=#3,
  coltitle=white,
  title={#2},
  fonttitle=\bfseries\footnotesize,
  boxrule=0.65pt,
  arc=1.2mm,
  left=1.7mm,
  right=1.7mm,
  top=1.2mm,
  bottom=1.2mm,
  before upper={\justifying\scriptsize\setlength{\parindent}{0pt}\setlength{\parskip}{2pt}},
  #1
}
\makeatletter
\newcommand{\draftwraptable}[3]{%
  \iflabreview@visible
    \begin{table}[ht]\centering\begin{minipage}{#2}#3\end{minipage}\end{table}%
  \else
    \begin{wraptable}{#1}{#2}#3\end{wraptable}%
  \fi}
\makeatother

\newcommand{\keeppositivevspace}{\RenewDocumentCommand{\vspace}{s m}{\ifdim\dimexpr##2\relax>0pt\vskip##2\relax\fi}}
\newcommand{\layoutvspace}[1]{}
\AtBeginEnvironment{table}{\keeppositivevspace}

\titleformat{\section}{\sffamily\Large\bfseries\color{boxheader}}{\thesection}{0.7em}{}
\titleformat{\subsection}{\sffamily\large\bfseries\color{boxheader}}{\thesubsection}{0.7em}{}
\titleformat{\subsubsection}{\sffamily\normalsize\bfseries\color{boxheader}}{\thesubsubsection}{0.7em}{}
\titlespacing*{\section}{0pt}{2.4ex plus .6ex minus .2ex}{1.0ex plus .2ex}
\titlespacing*{\subsection}{0pt}{2.0ex plus .5ex minus .2ex}{0.8ex plus .2ex}
\titlespacing*{\subsubsection}{0pt}{1.6ex plus .4ex minus .2ex}{0.6ex plus .2ex}
\usepackage{needspace}
\preto\section{\needspace{8\baselineskip}}
\preto\subsection{\needspace{5\baselineskip}}

\DeclareCaptionFont{navy}{\color{boxheader}}
\renewcommand{\headrulewidth}{0.4pt}
\renewcommand{\headrule}{{\color{arxrule}\hrule height\headrulewidth width\headwidth}}
\fancypagestyle{firstpage}{%
  \fancyhf{}%
  \fancyfoot[C]{\sffamily\footnotesize\color{arxmuted}\thepage}%
  \renewcommand{\headrulewidth}{0pt}}

\hypersetup{
  colorlinks=true,
  linkcolor=arxaccent,
  citecolor=arxaccent,
  urlcolor=arxaccent,
  pdftitle={Constitution-Guided Watermarking},
  pdfauthor={Toluwani Aremu, Samuele Poppi, Nils Lukas}
}
\renewcommand{\bibfont}{\small}
\begin{document}
\thispagestyle{firstpage}

% ------------------------------------------------------------ title block
\begin{flushleft}
{\sffamily\footnotesize\bfseries\color{arxaccent}PREPRINT\par}
\vspace{6pt}
{\sffamily\bfseries\color{boxheader}\fontsize{24}{29}\selectfont Constitution-Guided Watermarking\par}
\vspace{14pt}
{\large Toluwani Aremu\qquad Samuele Poppi\qquad Nils Lukas\par}
\vspace{4pt}
{\normalsize\color{arxmuted}MBZUAI, Abu Dhabi, United Arab Emirates\par}
\vspace{12pt}
{\color{boxheader}\rule{\linewidth}{1.1pt}\par}
\end{flushleft}
\vspace{2pt}

% --------------------------------------------------------------- abstract
\begin{tcolorbox}[
  enhanced,
  colback=arxpanel,
  frame hidden,
  sharp corners,
  borderline west={2.2pt}{0pt}{boxheader},
  left=12pt, right=12pt, top=9pt, bottom=9pt,
  fontupper=\small
]
{\sffamily\footnotesize\bfseries\color{boxheader}ABSTRACT\par}
\vspace{4pt}
Regulation such as the EU AI Act requires providers of language models to mark generated text in a machine-readable way.
Watermarking meets this requirement by embedding a secret-key signal during generation, so the provider can later detect its model's text.
A watermark should stay detectable, preserve quality and survive paraphrasing, but a stronger signal survives more paraphrasing and distorts more text.
Providers resolve this trade-off with one configuration for all requests, although a provider may want robustness for misuse-prone requests and no watermark for editing assistance, which the EU AI Act exempts.
We introduce \emph{constitution-guided watermarking}, in which the provider states these requirements as natural-language rules, a \emph{watermark constitution}, and each request receives its rule's configuration.
Offline, a reasoning agent reads the rules and the watermark code and refines each rule's configuration from measured detection, quality and robustness.
At deployment, a monitor routes each prompt to its rule's configuration without modifying the serving model.
Each configuration is stored with its evidence, so the provider can audit every deployment decision, including unmet requirements.
Our evaluation with the KGW watermark shows that (i) on misuse-prone requests, the framework detects more paraphrased text than fixed configurations and random search at matching clean detection and quality, (ii) it exempts all editing requests, and (iii) changing the stated priorities shifts the selected configuration as intended.

\end{tcolorbox}
\vspace{6pt}

% ------------------------------------------------------------------ body
% Figure 1. Rules and stored policies follow Table tab:scenarios and
% Table tab:selected-configs (reasoning verifier); both prompts are test-split
% prompts from code/results/predeployment/kgw_compact_20260908 (rules
% low_risk_editing and high_stakes_advice).
\begin{wrapfigure}{r}{0.50\textwidth}
\vspace{-1.2em}
\centering
\scriptsize
\setlength{\tabcolsep}{2.5pt}
\renewcommand{\arraystretch}{1.05}
\begin{minipage}{0.97\linewidth}
\centering
{\bfseries (a) Offline, one stored policy per rule}\\[2pt]
\begin{tabular}{@{}l l l@{}}
\toprule
\textbf{Rule} & \textbf{Policy} & \textbf{Status} \\
\midrule
Misuse-prone requests & KGW, bias 6 & best-effort \\
Low-risk editing      & no watermark & exempt \\
High-stakes advice    & KGW, bias 4 & passed \\
Deterministic answers & KGW, bias 4 & passed \\
General generation    & KGW, bias 4 & passed \\
\bottomrule
\end{tabular}\\[4pt]
{\bfseries (b) At deployment, the monitor routes each prompt}\\[2pt]
\begin{tabular}{@{}p{0.70\linewidth} l@{}}
\toprule
\textbf{Prompt} & \textbf{Rule, policy} \\
\midrule
``Rewrite the following medical disclaimer to improve readability: \ldots'' & Editing, none \\
\addlinespace[2pt]
``I have a persistent cough and shortness of breath; what should I do?'' & Advice, bias 4 \\
\bottomrule
\end{tabular}
\end{minipage}
\vspace{-0.6em}
\caption{\textbf{Two medical prompts receive different watermark policies.}
(a)~Offline, the watermark verification agent stores one KGW configuration per rule, where the bias sets the watermark's strength.
(b)~At deployment, the monitor routes each prompt, both from our test set, to its rule.}
\label{fig:wrapped}
\vspace{-1.em}
\end{wrapfigure}

\section{Introduction}
\label{sec:introduction}

Providers of language models need to identify text that their models generated~\citep{kirchenbauer2023watermark,synthid}.
Watermarking embeds a statistical signal into the generated tokens that only a holder of the secret key can detect.
Requests on one topic can still need different watermarks, as a provider may exempt a request to proofread a medical disclaimer, because the user wrote the text, yet watermark medical advice with as little distortion as possible.
Regulation draws the same line, as Article~50 of the EU AI Act requires machine-readable marking of synthetic content but exempts assistance with standard editing~\citep{eu_2024}.
Watermarking edited text also weakens attribution, as Anthropic notes that a detected watermark shows model involvement but does not distinguish model-authored text from human text that Claude heavily edited~\citep{anthropic2026claudewatermark}.
We call the configuration a provider applies to a request, or its decision not to watermark it, the request's \emph{watermark policy}.

Choosing a watermark policy is hard because the properties a provider wants conflict~\citep{zhao2024sok}.
For KGW, the watermark of \citet{kirchenbauer2023watermark} that adds a bias to the logits of a key-dependent subset of tokens, a larger bias makes the watermark easier to detect but distorts the text more~\citep{wouters2024optimizing}.
A watermark that survives benign edits also survives edits that change the meaning, which lets an attacker attribute harmful text to the provider~\citep{Pang2024NoFL}.
Fixed configurations recommended by benchmarks~\citep{Piet2023MarkMW,pan2024markllm} apply one operating point to every request, so quality-sensitive requests pay for robustness that only misuse-prone requests need.
Schemes designed around one property, such as distortion-free sampling~\citep{Kuditipudi2023RobustDW} or tournament sampling~\citep{synthid}, fix the trade-off when the scheme is designed.
Adaptive watermarks vary the signal with token uncertainty or generation features~\citep{liu2024adaptive,morphmark2025,wang2025learning}, but they optimize an objective that the designer fixes, not the one the provider assigns to each request.
This mismatch is an opportunity, as a system that knew which requirements apply to a request could choose the policy that meets them.

Providers can state such requirements as natural-language rules, such as ``do not watermark low-risk editing'' or ``for misuse-prone requests, prioritize robustness over quality''.
Finding a configuration that meets a rule requires generating, attacking and scoring watermarked text, but this search runs once per rule and before deployment.
Deployment then reduces to recognizing which rule a prompt triggers, as Figure~\ref{fig:wrapped} shows for the two medical prompts.
We introduce \emph{constitution-guided watermarking}, in which the provider writes these rules, the permitted watermark algorithms and \emph{gates}, which are thresholds on measured detection, quality and robustness, into a \emph{watermark constitution}.
As in Constitutional AI~\citep{bai2022constitutional}, rules written in natural language govern a system's behavior, but here they govern the watermark, and the serving model stays frozen.

Offline, a \emph{watermark verification agent} reads each rule and the watermark implementation and proposes a configuration, which a \emph{deterministic verifier} checks for legality before the configuration is measured on reference prompts.
The agent revises its proposals from these measurements until each rule passes its gates or the budget runs out, and stores the result as a \emph{certificate} marked reference-passed or best-effort.
At deployment, a \emph{deployment monitor} identifies the rule a prompt triggers and applies the configuration in that rule's certificate.
Our proof-of-concept instantiates both roles with pretrained models, \texttt{gpt-5.6-luna} as the agent and \texttt{gpt-5.4-nano} as the monitor.

We evaluate the framework with KGW, \texttt{Qwen3-32B} as the serving model and a five-rule constitution on 500 test prompts.
Against random search with the same evaluation budget, the agent's policies detect more paraphrased text on misuse-prone requests, 17.0\% versus 11.6\% after DIPPER and 13.8\% versus 12.2\% after DIAA.
Random search also needs ranking weights set by hand for each rule, which the agent replaces by reading the rule text.
Over the better of two fixed KGW configurations, detection after paraphrasing rises by 11 to 14 percentage points, while clean detection at a 0.1\% false-positive rate (97.5\% versus 92.5\%) and the quality pass rate (98.6\% versus 98.4\%) also improve.
The framework exempts all 100 editing requests, which both fixed configurations watermark.
Detection after paraphrasing stays below the provider's 50\% gate, so the framework exports the misuse rule as best-effort and records the unmet gate.
Policy execution also depends on the monitor, as replacing \texttt{gpt-5.4-nano} with \texttt{gpt-5-nano} lowers compliance with the rules' actions from 100\% to 75\%.

\textbf{Contributions.}
\begin{enumerate}[label=\arabic*.,leftmargin=*,itemsep=0pt,topsep=2pt]
  \item We propose constitution-guided watermarking, to our knowledge the first framework that turns provider-written natural-language rules into per-request watermark policies, each stored with its evidence, and that accepts any verifier proposing configurations under the same gates and budget.
  \item We show that a reasoning agent selects these policies from the rule text and the watermark implementation, without the per-rule ranking weights that random search requires.
  \item We construct constitution-labeled prompts through generation, critique and revision, and an evaluation that counts routing, generation and attack failures against the method.
\end{enumerate}
Appendix~\ref{sec:related-work} reviews related work.

\section{Problem Formalization}
\label{sec:formalization}

\textbf{Watermarking system.}
% We adopt the notation of \citet{zhao2024sok}, specializing to zero-bit text watermarks.
Let $\mathcal{M}$ be the frozen serving model, $\pi\in\Pi$ a prompt, and $x\in\mathcal{R}$ a response.
Unwatermarked generation is written $\mathcal{M}(\pi)\to x$.
For our framework, we also index each permitted algorithm $a\in\mathcal{W}$ by a legal configuration $\psi\in\Psi_a$:
\begin{equation}
    \mathsf{Watermark}_{\mathsf{gk}}^{\mathcal{M}}
    (\pi;a,\psi)\to x.
\end{equation}
Here, $\mathsf{gk}$ and $\mathsf{dtk}$ denote the generation and detection keys, which coincide for symmetric schemes such as KGW.
At nominal FPR $\alpha$, the configured detector is
\begin{equation}
    \mathsf{Detect}_{\mathsf{dtk}}^{\alpha}(x;a,\psi)
    =
    \begin{cases}
        \mathsf{true}, &
        s_{a,\psi,\mathsf{dtk}}(x)\geq
        \eta_{a,\psi,\mathsf{dtk},\alpha},\\
        \mathsf{false}, & \text{otherwise},
    \end{cases}
    \label{eq:detector}
\end{equation}
where $s$ is the detection score and $\eta$ is calibrated on null text.
We write $d_{a,\psi}^{\alpha}(x)$ for the indicator that $\mathsf{Detect}_{\mathsf{dtk}}^{\alpha}(x;a,\psi)=\mathsf{true}$.
% For compactness, define
% \[
%     d_{a,\psi}^{\alpha}(x)=
%     \mathbf{1}[\mathsf{Detect}_{\mathsf{dtk}}^{\alpha}
%     (x;a,\psi)=\mathsf{true}].
% \]
We fix the deployment key pair throughout the empirical evaluation.
Detection must match generation in its key, tokenizer, vocabulary size, and scoring-relevant parameters.
Generation-only parameters, such as KGW's logit bias $\delta$, can differ without changing the detector.
% Equation~\ref{eq:detector} assumes the configuration is known; Section~\ref{sec:method-recovery} considers detection without this metadata.

\textbf{Provider constitution.}
A constitution is
\begin{equation}
    \mathcal{C}=(\mathcal{W},\mathcal{S},\prec,\boldsymbol{\tau}),
\end{equation}
where $\mathcal{S}=\{r_1,\ldots,r_m\}$ is the rule set, $\prec$ specifies precedence, and $\boldsymbol{\tau}$ contains quantitative gates.
Each rule $r=(u_r,w_r,\mathbf{o}_r,\mathcal{G}_r)$ consists of its natural-language instruction $u_r$, which is authoritative, a watermarking decision $w_r\in\{\mathsf{WM},\mathsf{NoWM}\}$, ordered objectives $\mathbf{o}_r$, and active gates $\mathcal{G}_r$.
% This representation interprets the provider's instruction.

Let $\mathcal{S}(\pi)$ contain the rules applicable to $\pi$.
Precedence resolves conflicting watermarking decisions.
For compatible watermarked rules $\mathcal{S}_{\mathsf{WM}}(\pi)$, objective recurrence determines priority:
\begin{equation}
    n_j(\pi)=
    \sum_{r\in\mathcal{S}_{\mathsf{WM}}(\pi)}
    \mathbf{1}[j\in\mathbf{o}_r].
    \label{eq:recurrence}
\end{equation}
The combined profile orders objectives by decreasing $n_j(\pi)$, breaking ties by provider rule order and retaining the applicable gates.
Thus, two requests for quality outweigh one request for robustness in the objective order without removing the robustness requirement.

\textbf{Constitutional objectives.}
For a rule profile $c$, let $\mathcal{D}_c$ be its prompt distribution and $\mathcal{A}_{\mathrm{FPR}}$ the provider's nominal FPR regimes.
For each evaluated prompt $\pi_i$, generate
$\mathsf{Watermark}_{\mathsf{gk}}^{\mathcal{M}}(\pi_i;a,\psi)\to x_i$.
The \emph{empirical detectability rate} is
\begin{equation}
    \widehat{D}_c(a,\psi)
    =
    \min_{\alpha\in\mathcal{A}_{\mathrm{FPR}}}
    \frac{1}{n_c}\sum_{i=1}^{n_c}
    d_{a,\psi}^{\alpha}(x_i).
    \label{eq:detectability}
\end{equation}
For the fixed detection key, empirical FPR on a null set $\mathcal{D}_0$ disjoint from threshold calibration is
\begin{equation}
    \widehat{F}_{\alpha}(a,\psi)
    =
    \frac{1}{|\mathcal{D}_0|}
    \sum_{z\in\mathcal{D}_0}d_{a,\psi}^{\alpha}(z).
    \label{eq:fpr}
\end{equation}
% This distribution-specific estimate is distinct from the SoK's FPR guarantee over randomly sampled detection keys for arbitrary fixed content.
% If null measurements guide configuration search, they constitute reference feedback, not an independent final audit.
% For each $\alpha$, we report the observed criterion
% $\widehat{F}_{\alpha}\leq\alpha$ separately from
% $U_{1-\beta}(F_{\alpha})\leq\alpha$, where $U_{1-\beta}$ is a one-sided binomial confidence bound.
% Neither criterion follows from satisfying a looser global FPR cap, and even perfect observed detection does not establish a population guarantee.

For a reference language model $q$ and a global natural-text baseline $b_q$ fixed before search, the \emph{quality pass rate} is
\begin{equation}
    \widehat{Q}_c(a,\psi)
    =
    \frac{1}{n_c}\sum_{i=1}^{n_c}
    \mathbf{1}\!\left[\operatorname{PPL}_q(x_i)\leq b_q\right].
    \label{eq:quality}
\end{equation}
Low-distortion or exactness requirements also need task-specific measurements, such as
$L_c(x_i,x_i^0)$ relative to $\mathcal{M}(\pi_i)\to x_i^0$ or an appropriate task reference.
The PPL gate alone establishes neither task correctness nor distribution preservation.

\textbf{Adversaries.}
We consider two adversaries among the provider's users, neither of whom holds the secret key.
A \emph{removal adversary} receives a watermarked response and paraphrases it to evade detection, for example to present generated text as human-written.
A non-adaptive removal adversary uses a general-purpose paraphraser, whereas an adaptive one also knows the watermark algorithm and optimizes its paraphraser against it~\citep{diaa2024optimizing}.
The core security property against this adversary is \emph{robustness}, which states that a watermarked response remains detectable after each attack that its rule names.
Formally, let $\mathcal{E}$ be such an attack, which we call an attack channel.
Let $\mathcal{T}_c$ contain the channels relevant to profile $c$, and let
$x'_{i,b}\leftarrow\mathcal{E}(x_i)$ be $B$ independently randomized attacks conditional on each response.
At $\alpha_*=\min\mathcal{A}_{\mathrm{FPR}}$, \emph{empirical robustness} is
\begin{equation}
    \widehat{R}_c(a,\psi)
    =
    \min_{\mathcal{E}\in\mathcal{T}_c}
    \frac{1}{n_cB}\sum_{i=1}^{n_c}\sum_{b=1}^{B}
    d_{a,\psi}^{\alpha_*}(x'_{i,b}).
    \label{eq:robustness}
\end{equation}
We report attack validity and execution failures separately and cluster uncertainty estimates by source prompt.
Robustness is evaluated only when activated by the rule.
For $\mathsf{NoWM}$ decisions, responses instead follow $\mathcal{M}(\pi)\to x$ and are evaluated for action compliance and any applicable quality requirements.
End-to-end evaluation retains generation and routing failures in its attempted-prompt denominators and reports their counts separately.

% \textbf{Forgery resistance.}
A \emph{forgery adversary} instead seeks acceptance of content that the provider did not issue, such as harmful text attributed to the provider.
The corresponding property is \emph{forgery resistance}, which states that such content is accepted only with small probability.
For profile $c$, let $\mathcal{H}_c$ denote such target content and $\mathsf{Acc}_c$ the provider's acceptance rule.
An adversary $\mathcal{B}$ observing $N$ watermarked responses produces $x^\star$, with success probability
\begin{equation}
    \operatorname{Adv}^{\mathrm{forge}}_c(\mathcal{B},N)
    =
    \Pr\!\left[x^\star\in\mathcal{H}_c
    \ \land\ \mathsf{Acc}_c(x^\star)=1\right].
    \label{eq:forgery-success}
\end{equation}
A compatible extension randomly selects one of $K\geq2$ secret keys within each profile and accepts text only when exactly one key detects it~\citep{aremu2025randomizedkeys}.
For blind attackers unable to separate samples by key, assuming key-symmetric, mutually independent detection outcomes on forged text, success is bounded by $(1-1/K)^{K-1}$ per attempt, irrespective of $N$ while these assumptions hold.
Appendix~\ref{app:forgery} gives the construction, assumptions, proof, and FPR calibration.
% This is a conditional statistical defense for compatible watermarks, not cryptographic unforgeability or a consequence of constitutional adaptation alone.
% Our experiments do not empirically validate forgery resistance.
% ; forgery is therefore not reported as a passed empirical gate.

% \textbf{Security requirements.}
% For security parameter $\lambda$, use the SoK's oracle notation to define the keyless distinguishing advantage
% \begin{equation}
% \begin{aligned}
%     \operatorname{Adv}^{\mathrm{pub}}_{a,\psi}(D)
%     =\Big|&
%     \Pr[D^{\mathcal{M},\mathcal{M}}(1^\lambda)\to 1]
%     &-
%     \Pr_{\mathsf{gk}}
%     [D^{\mathcal{M},W_{a,\psi,\mathsf{gk}}}
%     (1^\lambda)\to 1]\Big|,
% \end{aligned}
%     \label{eq:public-undetectability}
% \end{equation}
% Here, $W_{a,\psi,\mathsf{gk}}$ abbreviates the configured watermarked generator.
% Undetectability requires negligible advantage for every efficient distinguisher $D$ with adaptive oracle access.
% We reserve $\mathsf{Attribute}_{\mathsf{ak}}$ for attribution: unforgeability is not merely successful detection.
% Our configuration study establishes neither property.
% Any claimed security guarantee must identify the supporting construction, adversary access, and assumptions.
% Changing hyperparameters or using multiple keys does not by itself establish such a guarantee.

\textbf{Feasibility and best effort.}
For gate $g\in\mathcal{G}_c$, let $\widehat{m}_{c,g}(a,\psi)$ be its measured value, $\tau_g$ its threshold, and $\bowtie_g$ its comparison.
Empirical acceptance requires
\begin{equation}
    \operatorname{Pass}_c(a,\psi)
    =
    \bigwedge_{g\in\mathcal{G}_c}
    \left[\widehat{m}_{c,g}(a,\psi)
    \ \bowtie_g\ \tau_g\right].
    \label{eq:constitution-pass}
\end{equation}
The provider's objective order $\mathbf{o}_c$ determines the incumbent among tested passing configurations.
If none passes within the search budget, the system returns the best measured incumbent under that order, labels it \textsc{Best-Effort}, and reports it explicitly.
% Objective priorities resolve trade-offs but never weaken hard gates.

\section{Conceptual Approach}
\label{sec:methodology}

Constitution-guided watermarking moves the costly search for configurations offline, so that deployment only routes each prompt to an evaluated configuration.
In step~(i) of Figure~\ref{fig:methodology-framework}, we generate prompts labeled with the rules they trigger, as Section~\ref{sec:method-data} describes.
In step~(ii), a watermark verification agent $A_\theta$ searches for one configuration per rule using measured feedback, as Section~\ref{sec:method-loop} describes.
In step~(iii), a deployment monitor $R_\varphi$ routes each prompt to the configuration of its rule, as Section~\ref{sec:method-deployment} describes.
Appendix~\ref{app:algorithms} gives pseudocode for steps~(ii) and~(iii), and Appendix~\ref{app:training-objectives} gives objectives for training both roles, which our proof-of-concept leaves pretrained.

% Figure 2. Vector rebuild of the former framework.png; box names match the
% wording of Section 4. Circled numbers match steps (i)-(iii) in the text.
\begin{figure}[t]
\centering
\newsavebox{\frameworkbox}
\sbox{\frameworkbox}{%
\begin{tikzpicture}[
  font=\scriptsize,
  >={Stealth[length=1.5mm,width=1.2mm]},
  box/.style={draw=#1!55!black, fill=#1!8, rounded corners=1.5pt, align=center,
    minimum height=10mm, inner sep=2pt, line width=0.45pt,
    execute at begin node={\hyphenpenalty=10000\exhyphenpenalty=10000}},
  box/.default=gray,
  arr/.style={->, draw=black!60, line width=0.55pt},
  frame/.style={draw=black!25, rounded corners=2pt, line width=0.45pt},
  framelabel/.style={font=\scriptsize\scshape, text=black!60, fill=white, inner sep=1.5pt},
  stepnum/.style={circle, draw=black!55, fill=white, inner sep=0.4pt, minimum size=3.4mm,
    font=\tiny\bfseries},
]
% Inputs
\node[box=gray, text width=40mm] (con) at (20mm,0)
  {\textbf{Provider constitution}\\rules, precedence and gates};
\node[box=gray, text width=40mm] (wm) at (65mm,0)
  {\textbf{Watermark algorithm}\\implementation and legal parameters};
\node[box=gray, text width=40mm] (data) at (110mm,0)
  {\textbf{Constitution-specific data}\\generate, critique and revise prompts};
\node[stepnum] at (data.north west) {i};

% Constitution-guided search
\node[box=teal, text width=31mm] (agent) at (16mm,-19mm)
  {\textbf{Watermark verification agent} $A_\theta$\\proposes a configuration per rule};
\node[box=teal, text width=26mm] (ver) at (49mm,-19mm)
  {\textbf{Deterministic verifier}\\rejects illegal configurations};
\node[box=orange, text width=26mm] (eval) at (80mm,-19mm)
  {\textbf{Evaluation}\\measures each gate on reference prompts};
\node[box=green, text width=26mm] (cert) at (111mm,-19mm)
  {\textbf{Certificate}\\\textsc{Reference-Passed} or \textsc{Best-Effort}};
\draw[arr] (agent) -- (ver);
\draw[arr] (ver) -- (eval);
\draw[arr] (eval) -- (cert);
\draw[arr, dashed, draw=teal!60!black] (eval.south) -- ++(0,-4mm) coordinate (fb)
  -| node[pos=0.25, below, font=\scriptsize, text=teal!50!black]
  {measurements and unmet gates, for a fixed number of rounds} (agent.south);

% Input arrows
\draw[arr] (con.south) -- (con.south |- agent.north);
\draw[arr] (wm.south) -- ++(0,-3mm) -| ([xshift=8mm]agent.north);
\draw[arr] (data.south) -- ++(0,-3mm) -| (eval.north);

% Constitution-guided deployment
\node[box=blue, text width=11mm] (prompt) at (5.5mm,-47mm) {\textbf{Prompt}\\$\pi$};
\node[box=blue, text width=26mm] (mon) at (30mm,-47mm)
  {\textbf{Deployment monitor} $R_\varphi$\\identifies the rule and retrieves its certificate};
\node[box=blue, text width=26mm] (run) at (62mm,-47mm)
  {\textbf{Trusted runtime}\\validates the certificate and resolves the key};
\node[box=blue, text width=26mm] (llm) at (94mm,-47mm)
  {\textbf{Serving model} $\mathcal{M}$\\generates with or without the watermark};
\node[box=green, text width=12.5mm] (resp) at (120.5mm,-47mm) {\textbf{Response}\\$x$};
\draw[arr] (prompt) -- (mon);
\draw[arr] (mon) -- (run);
\draw[arr] (run) -- (llm);
\draw[arr] (llm) -- (resp);
\draw[arr] (cert.south) -- ++(0,-12mm) -| (mon.north);

% Frames, drawn behind the boxes
\begin{scope}[on background layer]
  \node[frame, fit=(agent)(cert)(fb), inner xsep=2.5mm, inner ysep=2mm,
    yshift=-1.5mm] (search) {};
  \node[frame, fit=(prompt)(resp), inner xsep=2.5mm, inner ysep=2.5mm,
    yshift=-1.5mm] (deploy) {};
\end{scope}
\node[framelabel, anchor=east] at ([xshift=-4mm]search.south east)
  {Constitution-guided search};
\node[stepnum] at (search.north west) {ii};
\node[framelabel, anchor=east] at ([xshift=-4mm]deploy.south east)
  {Constitution-guided deployment};
\node[stepnum] at (deploy.north west) {iii};
\end{tikzpicture}%
}
\usebox{\frameworkbox}
\caption{\textbf{Constitution-guided watermarking searches for configurations offline, so deployment only routes prompts.}
(i)~Constitution-specific data supplies prompts labeled with the rules they trigger.
(ii)~The watermark verification agent proposes one configuration per rule, the deterministic verifier rejects illegal ones, and the evaluation measures each gate on reference prompts.
Measurements and unmet gates return to the agent, and the final configurations are stored in a certificate.
(iii)~At deployment, the monitor identifies the rule a prompt triggers, a trusted runtime validates that rule's certificate and resolves the secret key, and the frozen serving model generates the response with or without the watermark.}
\label{fig:methodology-framework}
\end{figure}

% \subsection{Training}
% \label{sec:method-training}

% \textbf{Watermark verification agent.} The agent learns to connect constitutional requirements to watermark mechanisms, parameter effects, and proof assumptions. 
% %
% Training examples pair implementations and measurement histories with legal proposals, source-supported explanations, and explicit trade-offs. 
% % Targets require expert
% % review or appropriate program and proof checkers; plausible explanations alone
% % are not valid supervision. Empirical comparisons can subsequently supply
% % preferences under the provider's objective order. 
% % Proof familiarity does not
% % make the agent a proof checker: unverified claims remain hypotheses.
% %
% \textbf{Deployment monitor.}
% Given a prompt $\pi$, the monitor learns to predict the applicable rule set $\widehat{\mathcal{S}}(\pi)$ and watermark action from constitution--prompt supervision, including overlaps. 
% % 
% % Provider precedence and
% % priority aggregation remain deterministic; supervision targets rule meaning,
% % not transient certificate identifiers. Held-out routing-development data
% % select confidence or abstention thresholds, with a provider-defined fallback
% % for uncertain requests. 
% % 
% We present both training objectives in details in Appendix~\ref{app:training-objectives}, and distinguish our pretrained instantiation in Section~\ref{sec:experiments}.

\subsection{Dataset Generation}
\label{sec:method-data}

Evaluating a constitution requires prompts labeled with the rules they trigger, which existing watermarking benchmarks do not provide.
We therefore construct constitution-specific data through a generate--critique--revise loop adapted from Constitutional AI~\citep{bai2022constitutional}.
Given the constitution and a target scenario, a \emph{generator} proposes a realistic user prompt.
An independent \emph{critic} then labels every applicable rule, the resolved watermark action, and the active gates, and rejects prompts that are unrealistic or do not match the target scenario.
A \emph{reviser} repairs rejected prompts, which are critiqued again; prompts that still fail are discarded.
We add paraphrases of accepted prompts, which test rule recognition beyond surface wording, and prompts that trigger several rules, which test it under overlapping requirements.
%
% Critic labels are treated as annotations rather than ground truth.
%
Before search begins, we fix a small rule-stratified reference set $\mathcal{D}_{\mathrm{ref}}$ and a disjoint test set $\mathcal{D}_{\mathrm{test}}$.
Each source prompt and its paraphrases are assigned to the same split to prevent leakage between them.
Only $\mathcal{D}_{\mathrm{ref}}$ enters the configuration loop; because it is reused across rounds, it serves as adaptation data, and all reported results use $\mathcal{D}_{\mathrm{test}}$.

\subsection{Constitution-Guided Search}
\label{sec:method-analysis}
\label{sec:method-loop}

The agent cannot tell from the rule text alone whether a configuration meets the rule's gates, so it refines its proposals from measurements on reference prompts.
At round $t$, the agent receives the natural-language rules, the watermark implementation and the reference measurements of earlier rounds.
The agent proposes one watermark policy $(a_c^{(t)},\psi_c^{(t)})$ per unresolved profile and states its rationale, assumptions, and expected effect on each active gate. 
The deterministic verifier then checks each proposal for source provenance, parameter legality, detector binding, profile coverage and novelty, and rejects proposals that fail.
%
% Source citations establish traceability, not the truth of a claimed guarantee. 
%
Reference answers follow $\mathsf{Watermark}_{\mathsf{gk}}^{\mathcal{M}} (\pi_i;a_c^{(t)},\psi_c^{(t)})\to x_i$ under each legal proposal.
For reasoning models, watermark generation and detection operate on the visible answer, not the preceding reasoning. 
We then calibrate detector thresholds using null text and use a separate null reference set to measure FPR. 
The evaluation measures only the profile's active gates, records failures and sample counts, and returns this evidence to the agent.
%
% Null-reference measurements also inform search and do not constitute an independent post-selection FPR audit.
% 
The agent retains the strongest measured incumbent under the profile's objective order and revises unsuccessful proposals.
When freezing is enabled, profiles satisfying all active gates are removed from subsequent search.
Unresolved profiles continue until the budget is exhausted, and their retained configurations are labeled best-effort.
After at most $T$ rounds, every profile has a passing record or an explicit \textit{best-effort} alternative. 
% 
% This is decision-level self-improvement. Specialist training can retain experience in $\theta$ across
% tasks, but the within-constitution loop does not require weight updates.

\subsection{Constitution-Guided Deployment}
\label{sec:method-certificate}
\label{sec:method-deployment}

Deployment must apply only configurations that the search evaluated, and must show the evidence behind each one.
The search therefore returns a versioned record that holds the constitution, the watermark implementation, the evaluation results and the selected configuration of each rule.
We call this record a certificate, and each rule's entry carries one of two statuses.
\textsc{Reference-Passed} means that observed reference estimates satisfy the gates, while \textsc{Best-Effort} exposes every unmet gate.
Exhausting the search budget establishes only that no passing configuration was found, not that the constitution is infeasible.
The provider decides whether to deploy a best-effort configuration.
For a prompt $\pi$, the monitor $R_\varphi$ predicts the applicable rules $\widehat{\mathcal{S}}(\pi)$ and retrieves their entries from the deployed policy $\Policy$.
For overlapping rules, provider precedence resolves the action and Equation~\ref{eq:recurrence} resolves compatible priorities. 
%
% A composite profile requires its own evaluated record; the monitor cannot claim that an atomic
% rule's certificate validates the combination. Missing or uncertain matches
% follow the declared fallback. Our compact experiment restricts configuration
% search to atomic rules.
% 
A trusted runtime then validates the retrieved entry and, when the rule requires a watermark, resolves the generation key $\mathsf{gk}$, so that neither the monitor nor the log sees a secret key.
The serving model $\mathcal{M}$ generates the response with the entry's configuration, or without a watermark for $\mathsf{NoWM}$ rules.

\section{Experiments}
\label{sec:experiments}
Our experiments address four questions, namely (i) whether constitution-guided watermarking improves on fixed and searched configurations, (ii) how feedback changes the agent's proposals, (iii) whether the selected policies follow the constitution's priorities and (iv) how the deployment monitor affects policy execution.

\textbf{Models.}
We instantiate the watermark verification agent $A_\theta$ with \texttt{gpt-5.6-luna} and the deployment monitor $R_\varphi$ with \texttt{gpt-5.4-nano}, both pretrained and at high reasoning effort.
The serving model is \texttt{Qwen3-32B}, which generates at most $8{,}192$ new tokens with at most $1{,}024$ in the visible answer, and we watermark only the visible answer, never the reasoning tokens.
We use KGW~\citep{kirchenbauer2023watermark} as implemented in MarkLLM~\citep{pan2024markllm}, and the agent may set the bias $\delta\geq1$, the green-list fraction $0<\gamma<1$, the context length $h\geq1$ and the seeding function $f\in\{\texttt{time},\texttt{additive},\texttt{skip},\texttt{min}\}$, where $\delta$ and $h$ are integers.
% Neither agent is
% trained here. These choices demonstrate the framework's operation, not the
% benefit of specialist training or calibrated applicability probabilities.
% Routing-development evidence is reported separately in Appendix~\ref{app:routing-pilot}.
% The key policy and remaining fields stay fixed. Every detector
% uses the evaluated configuration's binding and its own calibrated threshold.

\begin{wraptable}{r}{0.65\textwidth}
\vspace{-2em}
    \centering
    \caption{\textbf{Provider constitution}, in first-applicable precedence order.}
    \label{tab:scenarios}
    \small
    \setlength{\tabcolsep}{1.5pt} % Slightly increased padding for readability
    \renewcommand{\arraystretch}{1.10} % Better vertical spacing
    
    \begin{tabular}{lcr}
        \toprule
        \textbf{Scenario} & \textbf{Action} & \textbf{Constitutional priority} \\ 
        \midrule
        Misuse-prone requests & WM   & Robustness over quality \\
        Low-risk editing      & NoWM & No watermark application \\
        High-stakes advice    & WM   & Quality, low distortion, high detection \\
        Deterministic answers & WM   & Quality and detectability \\
        General generation    & WM   & Minimize unnecessary distortion \\
        \bottomrule
    \end{tabular}
    \vspace{-1em}
\end{wraptable}
\textbf{Constitution and data.}
The constitution has five rules, which Table~\ref{tab:scenarios} summarizes and Appendix~\ref{app:constitution} reproduces.
\texttt{GPT-OSS-120B} generates the prompts and \texttt{Qwen3-32B} critiques them.
The search uses 10 reference prompts per rule (50 in total) for at most five rounds, and the evaluation uses 100 disjoint test prompts per rule (500 per method), which never enter the search.

\textbf{Baselines.}
We compare with two fixed KGW configurations, Fixed KGW A with $(\gamma,\delta,h,f)=(0.25,4,3,\texttt{time})$ and Fixed KGW B with $(0.5,2,1,\texttt{time})$, which \citet{Piet2023MarkMW} and \citet{pan2024markllm} prespecify as general operating points rather than per-rule optima.
Within the framework, a random-search verifier replaces the agent, gets the same evaluation budget and monitor, so the two verifiers differ only in the configurations they store, which Appendix~\ref{app:selected-configs} lists.
% The three arms share prompts, paired seeds, decoding limits, and evaluators: $1{,}500$ test attempts in total.
%
% Separate unwatermarked and editing-exemption-matched fixed controls remain
% planned. The completed optimizer comparison evaluates the retained agent
% policy against random search under a maximum of five reference evaluations
% per watermarked rule, using the same 10 prompts. Both use annotated rule
% assignments, exempt editing, and regenerate answers on the same 500 test
% prompts. This removes learned routing from the optimizer comparison; it is
% not an output-preserving counterfactual of the Nano arm. The methods share
% an empirical evaluation budget, not a compute budget or an identical
% selection objective. The heuristic produced no evaluated policy, and Bayesian
% optimization was not run. Constitution sensitivity is ongoing and is excluded
% from the reported results. Appendix~\ref{app:followups} specifies these
% protocols and their limitations.

\textbf{Gates and attacks.}
Each watermarked rule requires (i) detection of at least 0.95, (ii) a quality pass rate of at least 0.90, (iii) robustness of at least 0.50 and (iv) an empirical FPR of at most 0.01.
We calibrate detectors at target FPRs $\alpha\in\{0.01,0.001\}$ on $5{,}000$ natural texts and measure the reference FPR on another $5{,}000$ null texts.
The quality pass rate uses the answer-only perplexity under \texttt{Qwen2.5-7B-Instruct} with the threshold $b_q=18.25$, the 95th percentile of 200 natural texts, fixed before the search.
For robustness, DIPPER~\citep{krishna2023paraphrasing}, a general-purpose paraphraser, and DIAA, the adaptive paraphraser of \citet{diaa2024optimizing} that is trained against KGW, each produce five attacks per answer.
We report the two attacks separately and gate their minimum.
Whether an attack preserves the answer's meaning requires a separate audit, which we do not perform.
Appendix~\ref{app:implementation} gives all decoding, budget, attack and baseline settings.
% Their feedback enters search, so they are not an untouched final audit. 
% Detection gates must hold at both regimes, but the shared 1\% FPR cap does not establish 0.1\% operation. The intended text-only follow-up uses fresh calibration and audit null sets; no valid completed recovery audit is reported here.
% No wrong-key or forgery-rate experiment is included.

\textbf{Metrics.}
Detection uses the detector bound to the generation record, which assumes that the configuration is known.
Appendix~\ref{sec:method-recovery} discusses this assumption and an unevaluated extension for prompt-free detector selection.
Clean detection covers the four watermarked rules ($n=400$), and the quality pass rate is the fraction of these prompts whose answer has a perplexity of at most $b_q$, which measures fluency.
Robustness is detection after each attack at a nominal 0.1\% FPR on the 100 misuse prompts, and exemption compliance is the fraction of editing requests answered without the watermark.
Routing, generation and attack failures count as failures in every denominator.
Intervals are 95\% bootstrap intervals that resample source prompts and keep each response's attacks together.
Appendix~\ref{app:fpr} reports the observed false positives on unwatermarked text at both FPR targets with exact binomial intervals.
These null texts also served as reference feedback during the search, so they do not form an independent audit, and passing the 1\% FPR gate does not establish a 0.1\% FPR.

\label{sec:results}

\begin{table}[t]
\centering
\small
\caption{\textbf{Constitutional-KGW raises robustness on misuse-prone requests and matches the best clean detection, quality and exemption compliance.}
% Constitutional-KGW (C-KGW) and random search use oracle rule assignments. 
% Fixed baselines watermark every request. 
Clean detection covers watermark-required rules ($n=400$); 
quality measures PPL $\le 18.25$ on the same prompts. 
Robustness covers 100 robustness-required prompts (5 attacks/response). 
Brackets show 95\% percentile-bootstrap intervals, clustered by prompt for robustness. 
The two constitutional rows share the framework, monitor, gates, and budget, and differ only in the verifier. All values are percentages; best results are \textbf{bolded}. Robustness at $\alpha=1\%$ is in Appendix~\ref{app:robust-1pct}.}
\label{tab:main-comparison}
\setlength{\tabcolsep}{3pt}
\renewcommand{\arraystretch}{1.18}
\resizebox{\linewidth}{!}{%
\begin{tabular}{@{} l cccccc @{}}
\toprule
& \multicolumn{2}{c}{\textbf{Clean Detection}\,$\uparrow$} 
& \textbf{Quality}\,$\uparrow$ 
& \multicolumn{2}{c}{\textbf{Robustness}\,$\uparrow$ ($\alpha = 0.1\%$)} 
& \textbf{Compliance}\,$\uparrow$ \\
\cmidrule(lr){2-3} \cmidrule(lr){4-4} \cmidrule(lr){5-6} \cmidrule(lr){7-7}
\textbf{Method} 
& $\alpha = 1.0\%$ 
& $\alpha = 0.1\%$ 
& PPL $\le 18.25$ 
& DIAA 
& DIPPER 
& Exempt \\
\midrule
\multicolumn{7}{@{}l}{\textit{Fixed baselines}} \\
\quad Fixed KGW A 
& $95.5\;_{[93.5,\,97.5]}$ 
& $92.5\;_{[90.0,\,95.0]}$ 
& $98.4\;_{[97.2,\,99.4]}$ 
& $\phantom{0}2.8\;_{[1.0,\,4.8]}$ 
& $\phantom{0}3.0\;_{[1.6,\,4.6]}$ 
& \phantom{00}0.0 \\
\addlinespace[2pt]
\quad Fixed KGW B 
& $80.0\;_{[76.3,\,83.8]}$ 
& $61.8\;_{[57.3,\,66.3]}$ 
& $98.2\;_{[97.0,\,99.2]}$ 
& $\phantom{0}1.4\;_{[0.4,\,2.4]}$ 
& $\phantom{0}8.6\;_{[6.0,\,11.4]}$ 
& \phantom{00}0.0 \\
\midrule
\multicolumn{7}{@{}l}{\textit{Constitutional framework}} \\
\quad Random-search verifier 
& $\mathbf{98.0}\;_{[96.5,\,99.3]}$ 
& $\mathbf{97.5}\;_{[96.0,\,99.0]}$ 
& $97.2\;_{[95.8,\,98.6]}$ 
& $12.2\;_{[8.0,\,17.0]}$ 
& $11.6\;_{[8.4,\,15.0]}$ 
& $\mathbf{100.0}$ \\
\addlinespace[2pt]
\quad \textbf{Reasoning verifier (C-KGW, ours)}
& $\mathbf{98.0}\;_{[96.5,\,99.3]}$ 
& $\mathbf{97.5}\;_{[96.0,\,99.0]}$ 
& $\mathbf{98.6}\;_{[97.4,\,99.6]}$ 
& $\mathbf{13.8}\;_{[10.8,\,16.8]}$ 
& $\mathbf{17.0}\;_{[12.8,\,21.4]}$ 
& $\mathbf{100.0}$ \\
\bottomrule
\end{tabular}%
}
\end{table}

\textbf{Comparison with fixed and searched configurations.}
We first ask whether per-rule policies improve on configurations that ignore the rules.
Table~\ref{tab:main-comparison} reports clean detection, the quality pass rate, robustness to each attack and exemption compliance, with the fixed configurations above the two verifiers of our framework.
\textbf{The agent's policies raise robustness where the constitution asks for it, at no loss in clean detection or quality.}
On the misuse rule, detection after paraphrasing rises by 11.0 percentage points after DIAA and 14.0 after DIPPER over Fixed KGW A.
The agent's policies also detect more paraphrased text than random search, and Appendix~\ref{app:robust-1pct} shows that the gain after DIPPER holds under a paired bootstrap test.
Random search needs ranking weights set by hand for each rule, whereas the agent reads the rule text.
Robustness still stays below the provider's 50\% gate, so the framework exports the misuse rule as best-effort.
Both verifiers exempt all 100 editing requests, which the fixed configurations watermark.
% Detailed per-rule breakdowns are provided in Table~\ref{tab:rule-outcomes}.

\begin{wraptable}[12]{r}{0.46\linewidth}
\vspace{-22pt}
\centering
\small
\caption{\textbf{Feedback doubles the bias and raises robustness at no cost in quality.}
$\gamma = 0.5$, $h = 4$ and $f = \texttt{min}$ are unchanged; detection at $\alpha = 0.1\%$.
Percentages, best in \textbf{bold}; full trace in Appendix~\ref{app:qualitative}.}
\label{tab:titan-refinement}
\setlength{\tabcolsep}{4pt}
\renewcommand{\arraystretch}{1.18}
\resizebox{\linewidth}{!}{%
\begin{tabular}{@{} l cc @{}}
\toprule
\textbf{Reference-set measure}
& \textbf{Initial}
& \textbf{Final} \\
\midrule
KGW bias, $\delta$
& 3
& 6 \\
\addlinespace[2pt]
Clean detection\,$\uparrow$ ($\alpha = 0.1\%$)
& \phantom{0}90.0
& \textbf{100.0} \\
\addlinespace[2pt]
Quality pass rate\,$\uparrow$
& \textbf{100.0}
& \textbf{100.0} \\
\addlinespace[2pt]
Worst-channel robustness\,$\uparrow$
& \phantom{00}0.0
& \phantom{0}\textbf{10.0} \\
\midrule
Policy status
& --
& Best-effort \\
\bottomrule
\end{tabular}%
}
\vspace{-15pt}
\end{wraptable}

\textbf{Iterative refinement.}
We next ask how feedback changes the agent's proposals, using the misuse rule as the example.
Table~\ref{tab:titan-refinement} compares its first and final proposals on the 10 reference prompts.
The first proposal, with bias $\delta=3$, failed every robustness check, and the agent responded by doubling the bias to $\delta=6$ while keeping $\gamma$, $h$ and $f$.
\textbf{Feedback raised robustness at no cost in quality, but not enough to pass the gate.}
Robustness rose from 0\% to 10\%, short of the 50\% gate, so the framework exported the policy as best-effort and recorded the unmet gate.
With 10 reference prompts, these numbers illustrate the mechanism rather than establish generalization.

\begin{table}[ht]
\vspace{-1.5em}
\centering
\small
\caption{\textbf{The selected bias follows the stated priority, from $\delta=2$ for quality-first to $\delta=6$ for robustness-first.}
Results are on 100 held-out prompts.
Each variant runs the same pipeline under a different constitution priority; all selected $\gamma = 0.5$.
Configurations are $(\delta, h, f)$.
Quality measures PPL $\le 18.25$; robustness uses five attacks per response at nominal $\alpha = 0.1\%$.
Quality and clean-detection brackets are 95\% Clopper--Pearson intervals; robustness brackets are 95\% bootstrap intervals clustered by prompt.
All values are percentages. Best results are \textbf{bolded}.}
\label{tab:constitution-sensitivity}
\setlength{\tabcolsep}{3pt}
\renewcommand{\arraystretch}{1.18}
\resizebox{\linewidth}{!}{%
\begin{tabular}{@{} l c ccccc @{}}
\toprule
& 
& \multicolumn{2}{c}{\textbf{Clean Detection}\,$\uparrow$} 
& \textbf{Quality}\,$\uparrow$ 
& \multicolumn{2}{c}{\textbf{Robustness}\,$\uparrow$ ($\alpha = 0.1\%$)} \\
\cmidrule(lr){3-4} \cmidrule(lr){5-5} \cmidrule(lr){6-7}
\textbf{Constitution} 
& $(\delta, h, f)$
& $\alpha = 1.0\%$ 
& $\alpha = 0.1\%$ 
& PPL $\le 18.25$ 
& DIAA 
& DIPPER \\
\midrule
Quality-first 
& $(2, 1, \texttt{time})$
& $\phantom{0}93.0\;_{[86.1,\,97.1]}$ 
& $91.0\;_{[83.6,\,95.8]}$ 
& $\mathbf{98.0}\;_{[93.0,\,99.8]}$ 
& $\phantom{0}4.6\;_{[2.4,\,7.6]}$ 
& $\phantom{0}5.6\;_{[3.0,\,8.4]}$ \\
\addlinespace[2pt]
Robustness-first 
& $(6, 2, \texttt{additive})$
& $\mathbf{100.0}\;_{[96.4,\,100.0]}$ 
& $\mathbf{98.0}\;_{[93.0,\,99.8]}$ 
& $91.0\;_{[83.6,\,95.8]}$ 
& $\mathbf{15.6}\;_{[12.5,\,19.1]}$ 
& $\mathbf{18.6}\;_{[15.3,\,22.3]}$ \\
\addlinespace[2pt]
Balanced 
& $(4, 1, \texttt{time})$
& $\phantom{0}98.0\;_{[93.0,\,99.8]}$ 
& $97.0\;_{[91.5,\,99.4]}$ 
& $94.0\;_{[87.4,\,97.8]}$ 
& $13.2\;_{[8.6,\,18.0]}$ 
& $17.0\;_{[13.2,\,21.4]}$ \\
\bottomrule
\end{tabular}%
}
\vspace{-1em}
\end{table}

\textbf{Sensitivity to constitutional priorities.}
We then ask whether the selected policies follow the constitution's stated priority.
We run the pipeline under three single-rule constitutions that differ only in their priority, and Table~\ref{tab:constitution-sensitivity} reports each selected configuration on 100 test prompts.
All three keep $\gamma=0.5$, and a larger bias $\delta$ means a stronger watermark.
\textbf{The bias follows the stated priority, but neighboring priorities overlap.}
The quality-first constitution selects $\delta=2$, the robustness-first one selects $\delta=6$ with a different context length and seeding function, and the balanced one falls between them on every metric.
Its intervals overlap with those of robustness-first throughout, so this sample size does not separate adjacent priorities, and no variant exceeds 19\% robustness after paraphrasing.
Appendix~\ref{app:sensitivity} gives the exact wording of the three constitutions.

\begin{table}[ht]
\layoutvspace{-1.2em}
\centering
\small
\caption{\textbf{A weaker monitor lowers rule assignment, detection, quality and policy compliance, with the same configurations.} 
Clean detection is averaged over watermark-required rules ($n=400$); robustness covers 100 prompts under paraphrase attack at $\alpha = 0.1\%$; quality measures PPL $\le 18.25$ across advice and deterministic requests ($n=200$). 
Rates are percentages; best results are \textbf{bolded}. Evasion probes and a routing rerun are in Appendix~\ref{app:routing}.}
\label{tab:router-comparison}
\setlength{\tabcolsep}{4pt}
\renewcommand{\arraystretch}{1.18}
\resizebox{\linewidth}{!}{%
\begin{tabular}{@{} l cccccccc @{}}
\toprule
& \textbf{Rule}\,$\uparrow$ 
& \multicolumn{2}{c}{\textbf{Clean Detection}\,$\uparrow$} 
& \textbf{Quality}\,$\uparrow$ 
& \multicolumn{2}{c}{\textbf{Robustness}\,$\uparrow$ ($\alpha = 0.1\%$)} 
& \textbf{Policy Action}\,$\uparrow$ 
& \textbf{Routing}\,$\downarrow$ \\
\cmidrule(lr){3-4} \cmidrule(lr){5-5} \cmidrule(lr){6-7} \cmidrule(lr){8-8} \cmidrule(lr){9-9}
\textbf{Router Policy} 
& \textbf{Assignment} 
& $\alpha = 1.0\%$ 
& $\alpha = 0.1\%$ 
& PPL $\le 18.25$ 
& DIAA 
& DIPPER 
& Compliance 
& Failures (/500) \\
\midrule
\texttt{gpt-5-nano} (low reasoning) 
& \phantom{0}$70.6$ 
& $87.5$ 
& $84.5$ 
& $91.0$ 
& $10.2$ 
& $16.4$ 
& \phantom{0}$75.0$ 
& 36 \\
\addlinespace[2pt]
\texttt{gpt-5.4-nano} (high reasoning) 
& $\textbf{100.0}$ 
& $\textbf{98.0}$ 
& $\textbf{97.5}$ 
& $\textbf{98.0}$ 
& $\textbf{13.8}$ 
& $\textbf{17.0}$
& $\textbf{100.0}$ 
& \phantom{0}$\textbf{0}$ \\
\bottomrule
\end{tabular}%
}
\layoutvspace{-1em}
\end{table}

\textbf{Deployment monitor.}
Finally, we ask how much policy execution depends on the monitor.
We replace \texttt{gpt-5.4-nano} at high reasoning effort with \texttt{gpt-5-nano} at low effort, keep the agent's policy and regenerate answers for all 500 test prompts, as Table~\ref{tab:router-comparison} shows.
\textbf{A weaker monitor lowers every metric that depends on the rule, although the configurations are unchanged.}
\texttt{gpt-5-nano} assigns the correct rule to 70.6\% of prompts, fails on 36 prompts because it returns invalid policy decisions, and lowers compliance with the rules' actions from 100\% to 75\%, mostly by watermarking editing requests.
Robustness changes less, from 13.8\% to 10.2\% after DIAA and from 17.0\% to 16.4\% after DIPPER.
Because each monitor's answers are regenerated, these differences compare deployments rather than identical generations.

\section{Discussion and Conclusion}
\label{sec:discussion}

\textbf{Why per-rule policies help.}
A single configuration accepts the same trade-off between robustness and distortion for every request.
A constitution instead lets the provider pay for robustness only on misuse-prone requests, keep low-distortion configurations elsewhere and skip the watermark for editing, which no single configuration can do.

\textbf{Cost.}
The search runs once per constitution and before deployment, so each request adds only one monitor call.
On the 500 test prompts, \texttt{gpt-5.4-nano} at high reasoning effort takes a median of 2.5\,s per call, 4.4\,s at the 95th percentile, and uses about $1{,}100$ input and 310 output tokens.

\textbf{Towards stronger robustness.}
Detection after paraphrasing reaches 13.8\% after DIAA and 17.0\% after DIPPER, below the provider's 50\% gate, so the framework exports the misuse rule as best-effort.
This gap reflects the tested policies and search budget rather than a ceiling for KGW. Appendix~\ref{app:hypotheses} lists changes that may close it, such as rule-specific starting points and other watermark families.

\textbf{Towards more watermark families and constitutions.}
Our evaluation covers KGW with one serving model, key pair and search seed, one five-rule constitution and three single-rule priority variants.
Repeating the search across seeds, watermark families and constitutions would show how far per-rule policies generalize.

\textbf{Larger reference sets.}
The search selects each policy on 10 reference prompts per rule that it reuses across rounds, so its measurements are finite-sample evidence rather than guarantees.
Larger reference sets, or fresh prompts in each round, would tighten these estimates.

\textbf{Task quality and attack validity.}
The quality pass rate measures fluency through perplexity and does not establish task correctness or low distortion, and we do not audit whether attacked outputs preserve the answer's meaning.
Task-specific quality measures and a semantic-validity audit of the attacks would close both gaps.

\textbf{Towards key-free detection and forgery resistance.}
Detection assumes access to the generation configuration, and we do not evaluate prompt-free detector selection or measure forgery resistance.
Appendix~\ref{sec:method-recovery} outlines prompt-free detection, and the randomized-key defense of \citet{aremu2025randomizedkeys} applies within each rule, as Appendix~\ref{app:forgery} shows.

\textbf{Stronger baselines.}
Besides fixed configurations and random search, a Bayesian-optimization verifier and hand-designed policies would further isolate the contribution of the reasoning verifier.
Appendix~\ref{app:limitations} discusses all limitations in more detail.

\textbf{Conclusion.}
We introduce constitution-guided watermarking, which turns provider-written natural-language rules into per-request watermark policies through an offline search and a deployment monitor, while the serving model stays frozen.
With KGW, its policies raise detection after paraphrasing on misuse-prone requests by up to 14 percentage points over a fixed configuration, match the best clean detection at a 0.1\% false-positive rate (97.5\%) and exempt all 100 editing requests.
Robustness still falls short of the provider's 50\% gate, which the framework records as an unmet requirement.
Because each policy is stored with its evidence, every deployment decision can be audited.
We encourage providers to state their watermarking requirements as rules and to evaluate a configuration for each rule against measured gates, rather than deploying one configuration for every request.

% \section{Conclusion}
% We introduced constitution-guided watermarking, which translates provider-written natural-language rules into request-dependent watermark policies through offline, feedback-driven configuration search and constrained deployment routing, while keeping the serving model frozen. On KGW, the resulting policies improve paraphrase robustness on robustness-prioritized requests, match the best clean detection and quality among the evaluated baselines, and exempt editing requests as specified. The framework also reports what it cannot achieve: robustness remains far below the provider's target, weaker monitors substantially degrade policy execution, and adjacent constitutional priorities are only partially distinguished. Stronger watermark families, specialist-trained agents, and independent audits are natural next steps.

\section*{Ethics Statement}

Watermark detection is statistical evidence, not conclusive proof of authorship
or misconduct. Deployment must account for mixed human--AI authorship, false
attribution, prompt privacy, and key security. Evaluating misuse-prone requests
does not justify bypassing the serving model's safety constraints; provenance
control is not a substitute for content safety.

\section*{AI Use Statement}

In this work, we used generative AI tools to generate the synthetic, constitution-labeled prompt datasets (Section \ref{sec:method-data}; GPT-OSS-120B and Qwen3-32B) and to implement parts of the experimental code. We have not used generative AI tools for formulating mathematical claims or writing proofs. We also used generative AI tools to draft and edit parts of the manuscript. We have reviewed all AI-assisted work; for example, generated prompts were filtered by an independent critic and spot-checked by the authors; AI-assisted code was tested and verified by the authors. We take responsibility for the final content of this work, including text, claims or artifacts produced with the aid of generative AI.
\section*{Reproducibility Statement}

Appendix~\ref{app:implementation} gives all decoding, budget, attack, and baseline settings, and Appendix~\ref{app:constitution} reproduces the evaluated constitution verbatim. Table~\ref{tab:selected-configs} lists every selected configuration, and Table~\ref{tab:reference-fpr} every deployed detector with its calibrated thresholds. Section~\ref{sec:method-data} describes how the constitution-labeled prompts were generated and split. Code, prompts, and run artifacts, including source hashes, agent proposals, rejections, and retained incumbents, will be released upon publication.

\bibliography{bibliography/reference}
\bibliographystyle{iclr2026_conference}

\clearpage
\appendix
\raggedbottom
\section{Related Work}
\label{sec:related-work}

\textbf{Statistical text watermarking.}
A watermark for language models must be detectable by the holder of a secret key while leaving the text useful.
Token-level watermarks change how tokens are sampled during decoding and detect the resulting statistical pattern with the key.
KGW~\citep{kirchenbauer2023watermark} splits the vocabulary into a key-dependent green list for each context and raises the logits of green tokens, and Unigram~\citep{zhao2024provable} uses one fixed green list to improve robustness.
EXP~\citep{aaronson2023watermarking} samples tokens with keyed randomness, and SynthID~\citep{synthid} selects tokens through a keyed tournament.
Distortion-free and cryptographic schemes leave the output distribution unchanged or make the watermark undetectable without the key, under stated assumptions~\citep{Kuditipudi2023RobustDW,christ2024undetectable}.
Toolkits such as MarkLLM~\citep{pan2024markllm} implement these schemes behind one interface but leave the choice of configuration to the provider.

\textbf{Watermark trade-offs and attacks.}
Choosing that configuration is hard because the properties of a watermark conflict, so no configuration suits every use.
Paraphrasing weakens token-level signals, especially when the attacker adapts to the known algorithm~\citep{krishna2023paraphrasing,diaa2024optimizing,cheng2025selfinformation}.
Other work estimates the watermark from its outputs or studies text falsely attributed to the provider~\citep{jovanovic2024watermarkstealing,gloaguen2024discovering}.
Randomized key selection bounds forgery under a multi-key threat model~\citep{aremu2025randomizedkeys}, and key-dependent structure that persists across outputs can support monitoring~\citep{aremu2026monitoringprimitive}.
How these properties trade off depends on the scheme, the model, the decoding policy and the output distribution~\citep{Pang2024NoFL,wouters2024optimizing,Piet2023MarkMW}, so the right trade-off differs across uses.

\textbf{Adaptive watermarking.}
Adaptive watermarks ease the quality side of this trade-off by placing the signal where it costs the least.
They change where or how strongly they embed the watermark based on token entropy, semantics or learned generation features~\citep{liu2024adaptive,morphmark2025,wang2025learning}.
They optimize one objective, fixed by the designer, within each output.
Our work instead lets the provider set the objective of each request in natural language and selects a configuration for each rule from measured evidence.

\textbf{Configuration search.}
Selecting a configuration from measured evidence is a search problem, since each candidate must be evaluated on generated text.
\citet{wouters2024optimizing} formalize the trade-off between detection and generation quality as an optimization problem, and benchmarks recommend fixed settings as general operating points~\citep{Piet2023MarkMW,pan2024markllm}.
Random search~\citep{bergstra2012random} and Bayesian optimization~\citep{jones1998efficient} are standard black-box alternatives, and our evaluation uses random search as a verifier with the same evaluation budget.

\textbf{Constitutions and feedback-driven agents.}
Instead of a black-box optimizer, our verifier is an agent guided by written rules and measured feedback, two ideas from work on steering language models.
Constitutional AI~\citep{bai2022constitutional} supervises a model with written principles, and agents such as Reflexion~\citep{shinn2023reflexion} and Self-Refine~\citep{madaan2023selfrefine} improve their outputs from feedback over repeated attempts, without updating weights.
Unlike that work, our rules govern the watermark rather than the model's behavior, and the serving model stays unchanged.
%
% Specialist training and feedback-driven configuration search are complementary. In either case, agent outputs remain proposals: parameter checks, reference measurements, and a deterministic enforcer define what the system may validate and execute.

% ======================================================================
\section{Experimental Details}
\label{app:implementation}

\subsection{Search and Deployment Procedures}
\label{app:algorithms}
Algorithm~\ref{alg:offline-verification} gives the constitution-guided search of Section~\ref{sec:method-loop}, and Algorithm~\ref{alg:online-deployment} gives the deployment of Section~\ref{sec:method-deployment}.

\begin{algorithm}[ht]
\caption{Source-grounded constitutional configuration search}
\label{alg:offline-verification}
\small
\renewcommand{\arraystretch}{1.06}
\begin{tabularx}{\linewidth}{@{}>{\raggedleft\arraybackslash}p{1.4em}X@{}}
& \textbf{Input:} $A_\theta$, constitution $\Constitution$, source $S$,
legal domains $\{\ConfigSpace_a\}$, frozen $\mathcal{M}$,
$\mathcal{D}_{\mathrm{ref}}$, budget $T$ \\
& \textbf{Output:} policy $\Policy$ with measured incumbents and gate status \\
\addlinespace[3pt]
1 & $U\gets$ watermarked profiles; $H_c\gets\emptyset$ for every $c\in U$ \\

& \quad {\footnotesize$\triangleright$ Refine configurations using reference-set feedback.} \\
2 & \textbf{for} $t=1,\ldots,T$ \textbf{while} $U\neq\emptyset$ \textbf{do} \\
3 & \quad $Z\gets A_\theta(\Constitution,S,U,\{H_c\}_{c\in U})$ \\

& \quad {\footnotesize$\triangleright$ Reject illegal configurations before generation.} \\
4 & \quad $Z\gets\textsc{ValidateOrRequestRepair}(Z,S,\{\ConfigSpace_a\})$ \\
5 & \quad $\widehat{\mathbf{m}}\gets
\textsc{Evaluate}(\mathcal{M},\mathcal{D}_{\mathrm{ref}},Z)$ \\
6 & \quad \textbf{for each} $c\in U$ \textbf{do} \\
7 & \qquad $H_c\gets H_c\cup\{(Z_c,\widehat{\mathbf{m}}_c)\}$ \\

& \qquad {\footnotesize$\triangleright$ Prefer passing candidates; otherwise retain the best measured trade-off.} \\
8 & \qquad $z_c^*\gets\textsc{BestMeasured}(H_c,\mathbf{o}_c,\mathcal{G}_c)$ \\
9 & \qquad \textbf{if} $\operatorname{Pass}_c(z_c^*)$ and freeze is enabled
\textbf{then} $U\gets U\setminus\{c\}$ \\
10 & \quad \textbf{end for} \\
11 & \textbf{end for} \\

& \quad {\footnotesize$\triangleright$ Attach evidence, exemptions, and explicit unmet requirements.} \\
12 & $\Policy\gets\textsc{BindEvidence}(\Constitution,S,\{z_c^*,H_c\})$ \\
13 & \textbf{return} $\Policy$ \\
\end{tabularx}
\end{algorithm}
\begin{algorithm}[t]
\caption{Constrained constitutional deployment}
\label{alg:online-deployment}
\small
\renewcommand{\arraystretch}{1.06}
\begin{tabularx}{\linewidth}{@{}>{\raggedleft\arraybackslash}p{1.4em}X@{}}
& \textbf{Input:} $R_\varphi$, prompt $\pi$, constitution $\Constitution$,
evidence index $\mathcal{I}$ for policy $\Policy$, frozen $\mathcal{M}$ \\
\addlinespace[3pt]

& \quad {\footnotesize$\triangleright$ Identify applicable rules and resolve overlapping requirements.} \\
1 & $\widehat{\mathcal{S}}(\pi)\gets
R_\varphi(\pi,\Constitution,\mathcal{I})$ \\
2 & $c\gets\textsc{ResolveAndAggregate}
(\widehat{\mathcal{S}}(\pi),\Constitution)$ \\
3 & $z\gets\textsc{RetrieveRecord}(c,\mathcal{I})$ \\

& \quad {\footnotesize$\triangleright$ Use the provider's fallback when no authorized policy can be applied.} \\
4 & \textbf{if} uncertain, missing, or unauthorized \textbf{then return}
\textsc{ProviderFallback}$(\pi,c)$ \\

& \quad {\footnotesize$\triangleright$ Enforce the retrieved action and its bound configuration.} \\
5 & $(w,a,\psi,\kappa)\gets\textsc{EnforceAndBind}(z,\mathcal{I})$ \\
6 & \textbf{if} $w=\mathsf{NoWM}$ \textbf{then} $\mathcal{M}(\pi)\to x$ \\

& \quad {\footnotesize$\triangleright$ Resolve secret keys only within the trusted runtime.} \\
7 & \textbf{else} $\mathsf{gk}\gets\textsc{ResolveGenerationKey}(\kappa)$ \\
8 & \quad $\mathsf{Watermark}_{\mathsf{gk}}^{\mathcal{M}}(\pi;a,\psi)\to x$ \\
9 & \textbf{end if} \\

& \quad {\footnotesize$\triangleright$ Record the executed policy without exposing secret keys.} \\
10 & \textsc{Audit}$(c,z,w,a,\psi,\kappa)$; \textbf{return} $x$ \\
\end{tabularx}
\par\smallskip
{\footnotesize $\kappa$ is an opaque key identifier. The trusted runtime
resolves $\mathsf{gk}$; the corresponding $\mathsf{dtk}$ is bound to the
record for detection. Neither the monitor nor the log receives secret keys.}
\end{algorithm}

\subsection{Implementation}
\label{app:impl-settings}
The main run uses seed $20260908$ and paired seeds for each prompt.
Watermarking and detection act only on the visible answer, never on the reasoning tokens.
The serving model reasons with temperature $0.6$, top-$p=0.95$ and top-$k=20$.
The agent may set only the fields that MarkLLM exposes in its KGW configuration file \texttt{KGW.json}, and the provider key, the left-context window policy and the calibrated detector operating points stay fixed.
The agent has a completion budget of $32{,}768$ tokens and the monitor one of $4{,}096$ tokens, and model identifiers are reported exactly as the API serves them.
The run artifacts keep source hashes, exact requests, rejected proposals, retained configurations and decoding settings.

\subsection{Attack Settings}
\label{app:attacks}
DIAA uses the \texttt{DDiaa/WM-Removal-KGW-Llama-3.1-8B} adapter on \texttt{Llama-3.1-8B-Instruct} with temperature $1$ and at most $1{,}024$ new tokens.
DIPPER-XXL uses lexical diversity $60$, order diversity $0$, a sentence interval of $1$, top-$p=0.75$ and at most $100$ new tokens per generation call.
Both run in bfloat16 and attack each answer five times.
Because the generation caps can truncate a paraphrase, detection after an attack measures how much of the signal survives, not whether the attack kept the answer's meaning, which we do not audit.

\subsection{Baselines}
\label{app:baselines}
\textbf{Fixed configurations.}
Fixed KGW A and B use $(\gamma,\delta,h,f)=(0.25,4,3,\texttt{time})$ and $(0.5,2,1,\texttt{time})$, which \citet{Piet2023MarkMW} and \citet{pan2024markllm} prespecify as operating points.
Both watermark every request, including editing.

\textbf{Budget-matched random-search verifier.}
The random-search verifier~\citep{bergstra2012random} replaces the agent in the same pipeline and receives the same evaluation budget, at most five legal candidates per watermarked rule, each measured on the same 10 reference prompts against the same active gates.
It samples $\gamma\sim\operatorname{Beta}(2,2)$, $\delta\sim\operatorname{Geom}(0.25)$ and $h\sim\operatorname{Geom}(0.5)$ on the positive integers, and $f$ uniformly from its legal values, which is an explicit search prior rather than a bounded grid.
Because it cannot read the rules, it ranks candidates first by whether they pass all active gates, then by a weighted, normalized shortfall from the gates, with attainment breaking ties.
The weights, set by hand for each rule, favor robustness for misuse, quality for advice and deterministic answers, and weight all gates equally for general generation.
The agent instead keeps the configuration that best fits each rule's stated priorities, judged from the rule text.
No test result selects a prior, weight, candidate or retained configuration.

At deployment, both verifiers share the deployment monitor, \texttt{gpt-5.4-nano} at high reasoning effort, and exempt editing as the constitution specifies, so they differ only in the configurations they store, which Table~\ref{tab:selected-configs} lists.
Answers are regenerated for each verifier on the same 500 test prompts.
The two verifiers share an evaluation budget, not a compute budget, and each uses one search seed.

\subsection{Provider Constitution}
\label{app:constitution}
The evaluated constitution contains the five rules below, reproduced verbatim in precedence order.
Precedence is first-applicable, so when several rules apply to a prompt, the monitor selects the first one listed, and only that rule's action and priorities apply.
Our experiments therefore do not exercise the objective recurrence of Equation~\ref{eq:recurrence}, which applies when a provider combines compatible rules.
Prompts that trigger several rules instead test whether the monitor applies precedence correctly.
The sensitivity constitutions of Appendix~\ref{app:sensitivity} each contain a single rule, so overlap resolution has no effect there.

\begin{tcolorbox}[
  enhanced,
  colback=white,
  colframe=boxheader,
  boxrule=0.7pt,
  arc=2.5mm,
  attach boxed title to top left={yshift*=-3mm, xshift=4mm},
  boxed title style={
    colback=boxheader,
    colframe=boxheader,
    sharp corners=downhill,
    arc=1.5mm,
    boxrule=0pt,
    top=1.5pt,
    bottom=1.5pt,
    left=6pt,
    right=6pt
  },
  title={\textsc{\textbf{\small System Policy Rules}} \normalfont\footnotesize(first-applicable)},
  top=8pt,
  bottom=6pt,
  left=8pt,
  right=8pt,
  shadow={0.5mm}{-0.5mm}{0mm}{black!10}
]
\small
\begin{description}[
  leftmargin=2.2cm,
  labelwidth=2.0cm,
  labelsep=0.2cm,
  itemsep=3.5pt,
  topsep=4pt,
  font=\normalfont\bfseries\color{boxheader}
]
  \item[Misuse.] For misuse-prone, malicious, or abusive requests, watermark the response and prioritize robustness over quality.
  \item[Editing.] Do not watermark grammar correction, proofreading, or low-risk text editing.
  \item[Advice.] For medical and financial advice, watermark the response and prioritize quality and low distortion while keeping detectability high.
  \item[Deterministic.] For deterministic answers, watermark the response and prioritize quality and detectability.
  \item[General.] Watermark all other responses while minimizing unnecessary distortion.
\end{description}
\end{tcolorbox}

% ======================================================================
\subsection{Sensitivity Constitutions}
\label{app:sensitivity}
The sensitivity study of Table~\ref{tab:constitution-sensitivity} reuses the general-generation slice, with the same 10 reference and 100 test prompts, serving model, KGW implementation, attacks, seeds, and gates for detectability, quality and robustness.
Only the natural-language priority instruction changes, as the three rules below show.

\begin{tcolorbox}[
  enhanced,
  colback=white,
  colframe=boxheader,
  boxrule=0.7pt,
  arc=2.5mm,
  attach boxed title to top left={yshift*=-3mm, xshift=4mm},
  boxed title style={
    colback=boxheader,
    colframe=boxheader,
    sharp corners=downhill,
    arc=1.5mm,
    boxrule=0pt,
    top=1.5pt,
    bottom=1.5pt,
    left=6pt,
    right=6pt
  },
  title={\textsc{\textbf{\small Sensitivity Policy Rules}}},
  top=8pt,
  bottom=6pt,
  left=8pt,
  right=8pt,
  shadow={0.5mm}{-0.5mm}{0mm}{black!10}
]
\small
\begin{description}[
  leftmargin=2.2cm,
  labelwidth=2.0cm,
  labelsep=0.2cm,
  itemsep=3.5pt,
  topsep=4pt,
  font=\normalfont\bfseries\color{boxheader}
]
  \item[Quality first.] Watermark these general-generation responses. Require detectability, quality, and robustness; prioritize quality over robustness.
  \item[Robustness first.] Watermark these general-generation responses. Require detectability, quality, and robustness; prioritize robustness over quality.
  \item[Balanced.] Watermark these general-generation responses. Require detectability, quality, and robustness; balance these three objectives equally.
\end{description}
\end{tcolorbox}

Each constitution runs the same agent for at most five rounds and keeps a passing or best-effort configuration.

\section{Additional Results}
\label{app:additional-results}

\subsection{Selected Per-Rule Configurations}
\label{app:selected-configs}
To see where the gains of Table~\ref{tab:main-comparison} come from, Table~\ref{tab:selected-configs} lists the configuration each verifier stores for every rule.
\textbf{The agent separates the misuse rule from the quality rules rather than tuning each rule.}
It gives the misuse rule its strongest bias and a longer context, and uses one configuration, $(0.5,4,1,\texttt{additive})$, for all three quality-prioritized rules.
Random search stores legal configurations that the rules' priorities do not motivate, such as $\delta=27$ for deterministic answers, yet matches the agent's clean detection.
The agent's advantage therefore lies mainly in robustness to paraphrasing on the misuse rule, which the next comparison tests at a looser FPR and with paired differences.

\begin{table}[ht]
\centering
\small
\caption{\textbf{Configurations stored per rule.}
Tuples are $(\gamma,\delta,h,f)$: green-list fraction, logit bias, context length, and seeding function.
Rules are listed in precedence order.
Reasoning-verifier configurations for advice, deterministic answers, and general generation are \textsc{Reference-Passed}; misuse is \textsc{Best-Effort} (Table~\ref{tab:titan-refinement}).
Random-search $\gamma$ is rounded for display only.}
% Both procedures exempt editing, as the constitution specifies.}
\label{tab:selected-configs}
\setlength{\tabcolsep}{6pt}
\renewcommand{\arraystretch}{1.10}
\begin{tabular}{@{} l l l @{}}
\toprule
\textbf{Rule (priority)} & \textbf{Reasoning verifier (C-KGW)} & \textbf{Random-search verifier} \\
\midrule
Misuse (robustness over quality)     & $(0.5, 6, 4, \texttt{min})$      & $(0.4109, 4, 2, \texttt{min})$ \\
Editing (exempt)                     & \textsc{NoWM}                    & \textsc{NoWM} \\
Advice (quality, low distortion)     & $(0.5, 4, 1, \texttt{additive})$ & $(0.5603, 5, 6, \texttt{time})$ \\
Deterministic (quality, detection)   & $(0.5, 4, 1, \texttt{additive})$ & $(0.6056, 27, 1, \texttt{skip})$ \\
General (minimize distortion)        & $(0.5, 4, 1, \texttt{additive})$ & $(0.5763, 6, 6, \texttt{min})$ \\
\bottomrule
\end{tabular}
\end{table}

\subsection{Robustness at Nominal 1\% FPR and Paired Comparison}
\label{app:robust-1pct}
Table~\ref{tab:robustness-1pct} repeats the robustness comparison at a nominal 1\% FPR.
\textbf{The ranking holds at this looser operating point.}
The agent's policies reach 27.8\% after DIAA and 30.8\% after DIPPER, ahead of random search and the fixed configurations, with gains of up to 19.2 percentage points over Fixed KGW A.
Robustness still stays below the provider's 50\% gate, which Equation~\ref{eq:robustness} defines at the strictest operating point, $\alpha_*=0.1\%$.

\begin{table}[ht]
\centering
\small
\caption{\textbf{At nominal $\alpha = 1\%$, Constitutional-KGW again has the highest robustness to both attacks.}
Post-paraphrase detection on the 100 robustness-required prompts (5 attacks/response), under the same generations, attacks, and routing as Table~\ref{tab:main-comparison}.
Brackets show 95\% percentile-bootstrap intervals, clustered by prompt.
All values are percentages. Best results are \textbf{bolded}.}
\label{tab:robustness-1pct}
\vspace{2pt}
\setlength{\tabcolsep}{14pt}
\renewcommand{\arraystretch}{1.18}
\begin{tabular}{@{} l cc @{}}
\toprule
& \multicolumn{2}{c}{\textbf{Robustness}\,$\uparrow$ ($\alpha = 1.0\%$)} \\
\cmidrule(lr){2-3}
\textbf{Method} & DIAA & DIPPER \\
\midrule
Fixed KGW A   & $\phantom{0}8.6\;_{[5.6,\,12.2]}$ & $11.8\;_{[8.6,\,15.4]}$ \\
Fixed KGW B   & $\phantom{0}8.8\;_{[5.8,\,12.0]}$ & $19.0\;_{[14.8,\,23.4]}$ \\
\midrule
Random-search verifier & $26.6\;_{[20.4,\,33.0]}$ & $28.2\;_{[23.0,\,33.6]}$ \\
\textbf{Reasoning verifier (C-KGW, ours)} & $\mathbf{27.8}\;_{[21.8,\,34.4]}$ & $\mathbf{30.8}\;_{[25.6,\,36.2]}$ \\
\bottomrule
\end{tabular}
\end{table}

Table~\ref{tab:paired-robustness} tests the difference between the agent and random search directly, with paired differences on the same prompts.
Every difference favors the agent, but only the gain of 5.4 percentage points after DIPPER at a nominal 0.1\% FPR excludes zero, and the other differences lie within sampling uncertainty at this sample size.

\begin{table}[ht]
\centering
\small
\caption{\textbf{The reasoning verifier's gain over random search excludes zero only after DIPPER at nominal $\alpha = 0.1\%$.} Paired robustness differences, reasoning verifier (C-KGW) minus random-search verifier.
Both methods are evaluated on the same 100 robustness-required prompts (5 attacks/response).
Brackets show 95\% paired percentile-bootstrap intervals over source prompts (10{,}000 replicates), keeping each prompt's attack repetitions together; attack failures count as misses.
Rates are percentages; differences are in percentage points.}
\label{tab:paired-robustness}
\vspace{2pt}
\setlength{\tabcolsep}{10pt}
\renewcommand{\arraystretch}{1.18}
\begin{tabular}{@{} c l ccc @{}}
\toprule
\textbf{Nominal FPR} & \textbf{Attack} & \textbf{Reasoning}\,$\uparrow$ & \textbf{Random search}\,$\uparrow$ & \textbf{Difference}\,$\uparrow$ \\
\midrule
$0.1\%$ & DIAA   & 13.8 & 12.2 & $+1.6\;_{[-2.6,\,3.8]}$ \\
        & DIPPER & 17.0 & 11.6 & $\mathbf{+5.4}\;_{[0.6,\,10.2]}$ \\
\addlinespace[2pt]
$1.0\%$ & DIAA   & 27.8 & 26.6 & $+1.2\;_{[-7.8,\,5.2]}$ \\
        & DIPPER & 30.8 & 28.2 & $+2.6\;_{[-3.8,\,9.2]}$ \\
\bottomrule
\end{tabular}
\end{table}

\subsection{Null-Reference False-Positive Rates}
\label{app:fpr}
\begin{table}[ht]
\centering
\small
\caption{\textbf{Measured null-reference FPRs, not an independent final audit.}
Each row tests 5,000 unwatermarked texts disjoint from threshold calibration; their feedback was reused during configuration selection.
$\eta_\alpha$ is the calibrated detection threshold at nominal FPR $\alpha$.
Brackets give two-sided 95\% Clopper--Pearson intervals; these are descriptive, not guarantees after adaptive selection.
The provider's looser 1\% cap is a separate requirement.
Detector tuples are $(\gamma, h, f)$ with fixed key and tokenizer; generation bias $\delta$ does not affect the score.
Random-search $\gamma$ is rounded for display only.
The reasoning verifier's three non-misuse rules share one detector, which coincides with Fixed KGW B's because for $h=1$ every seeding function reduces to the preceding token.
Rates are percentages.}
\label{tab:reference-fpr}
\setlength{\tabcolsep}{12pt}
\renewcommand{\arraystretch}{1.05}
\resizebox{\linewidth}{!}{%
\begin{tabular}{@{} l l cc cc @{}}
\toprule
& 
& \multicolumn{2}{c}{\textbf{Nominal} $\alpha = 1.0\%$}
& \multicolumn{2}{c}{\textbf{Nominal} $\alpha = 0.1\%$} \\
\cmidrule(lr){3-4} \cmidrule(lr){5-6}
\textbf{Detector / Use}
& $(\gamma, h, f)$
& $\eta_\alpha$
& Observed FPR\,$\downarrow$
& $\eta_\alpha$
& Observed FPR\,$\downarrow$ \\
\midrule
Fixed KGW A
& $(0.25, 3, \texttt{time})$
& 2.531 & $0.940\;_{[0.691,\,1.248]}$
& 3.408 & $0.080\;_{[0.022,\,0.205]}$ \\
\addlinespace[2pt]
Fixed KGW B
& $(0.5, 1, \texttt{time})$
& 2.776 & $1.060\;_{[0.795,\,1.384]}$
& 3.842 & $0.160\;_{[0.069,\,0.315]}$ \\
\midrule
Reasoning: non-misuse
& $(0.5, 1, \texttt{additive})$
& 2.776 & $1.060\;_{[0.795,\,1.384]}$
& 3.842 & $0.160\;_{[0.069,\,0.315]}$ \\
\addlinespace[2pt]
Reasoning: misuse
& $(0.5, 4, \texttt{min})$
& 2.371 & $1.040\;_{[0.778,\,1.362]}$
& 3.208 & $0.140\;_{[0.056,\,0.288]}$ \\
\midrule
Random search: deterministic
& $(0.6056, 1, \texttt{skip})$
& 2.967 & $1.000\;_{[0.743,\,1.316]}$
& 4.111 & $0.140\;_{[0.056,\,0.288]}$ \\
\addlinespace[2pt]
Random search: general
& $(0.5763, 6, \texttt{min})$
& 2.529 & $1.020\;_{[0.760,\,1.339]}$
& 3.475 & $0.080\;_{[0.022,\,0.205]}$ \\
\addlinespace[2pt]
Random search: advice
& $(0.5603, 6, \texttt{time})$
& 2.402 & $0.880\;_{[0.640,\,1.180]}$
& 2.986 & $0.180\;_{[0.082,\,0.341]}$ \\
\addlinespace[2pt]
Random search: misuse
& $(0.4109, 2, \texttt{min})$
& 2.383 & $1.140\;_{[0.865,\,1.475]}$
& 3.448 & $0.080\;_{[0.022,\,0.205]}$ \\
\bottomrule
\end{tabular}%
}
\end{table}
All of these detection rates assume that each detector holds its nominal FPR.
We therefore evaluate each deployed detector on 5{,}000 unwatermarked texts held out from threshold calibration, as Table~\ref{tab:reference-fpr} shows.
\textbf{Observed FPRs stay close to their nominal targets.}
They range from 0.88\% to 1.14\% at a nominal 1\% and from 0.08\% to 0.18\% at a nominal 0.1\%.
Several detectors slightly exceed their targets, although every 95\% interval includes the nominal rate.
Because these texts also informed configuration selection, we read the intervals descriptively, and an independent assessment of the selected policies would require a fresh set of unwatermarked text.

% ----------------------------------------------------------------------
% Styles for the qualitative examples (Appendix C)
% ----------------------------------------------------------------------
\definecolor{qpass}{RGB}{38,120,74}
\definecolor{qfail}{RGB}{176,52,44}
\definecolor{qwarn}{RGB}{168,106,18}
\definecolor{qmuted}{RGB}{104,112,124}
\definecolor{qpanel}{RGB}{243,245,248}
\definecolor{qline}{RGB}{206,211,219}

\newcommand{\qyes}{\textcolor{qpass}{\ding{51}}}
\newcommand{\qno}{\textcolor{qfail}{\ding{55}}}
\newtcbox{\qchip}[1][]{on line, enhanced, colback=white, colframe=qline,
  boxrule=0.5pt, arc=1.2mm, boxsep=0pt, left=3.5pt, right=3.5pt,
  top=1.6pt, bottom=1.6pt, fontupper=\footnotesize, #1}
\newcommand{\qrule}[1]{\qchip[colback=qpanel, colframe=qline]{\textsc{#1}}}
\newcommand{\qwm}{\qchip[colback=boxheader, colframe=boxheader,
  fontupper=\footnotesize\bfseries\color{white}]{WM}}
\newcommand{\qnowm}{\qchip[colframe=boxheader,
  fontupper=\footnotesize\bfseries\color{boxheader}]{NoWM}}
\newcommand{\qbesteffort}{\qchip[colback=qwarn!8, colframe=qwarn!70,
  fontupper=\footnotesize\bfseries\color{qwarn}]{\textsc{Best-Effort}}}
\newcommand{\qcorrect}{\qchip[colback=qpass!7, colframe=qpass!60,
  fontupper=\footnotesize\bfseries\color{qpass}]{\ding{51}\,Correct}}
\newcommand{\qwrong}{\qchip[colback=qfail!7, colframe=qfail!60,
  fontupper=\footnotesize\bfseries\color{qfail}]{\ding{55}\,Misrouted}}
\newcommand{\qstage}[1]{{\leavevmode\footnotesize\bfseries\scshape\color{boxheader}#1}}
\newcommand{\qfield}[1]{{\leavevmode\footnotesize\color{qmuted}#1}}
\newcommand{\qsay}[1]{\itshape``#1''}
\newcommand{\qgate}[3]{#1~\textbf{#2}\,#3}
\newcommand{\qdot}{\enspace{\color{qline}$\bullet$}\enspace}
\newcommand{\qdivider}[1]{\multicolumn{3}{@{}c@{}}{%
  {\color{qline}\rule[0.6ex]{0.28\linewidth}{0.4pt}}\enspace
  {\footnotesize\itshape\color{qmuted}#1}\enspace
  {\color{qline}\rule[0.6ex]{0.28\linewidth}{0.4pt}}}}
\newcommand{\qsep}{\multicolumn{2}{@{}c@{}}{%
  {\color{qline}\rule[0.5ex]{\linewidth}{0.4pt}}}}

\newtcolorbox{qbox}[1]{
  enhanced,
  colback=white,
  colframe=boxheader,
  boxrule=0.7pt,
  arc=2.5mm,
  attach boxed title to top left={yshift*=-3mm, xshift=4mm},
  boxed title style={colback=boxheader, colframe=boxheader,
    sharp corners=downhill, arc=1.5mm, boxrule=0pt,
    top=1.5pt, bottom=1.5pt, left=6pt, right=6pt},
  title={#1},
  top=10pt, bottom=8pt, left=10pt, right=10pt,
  shadow={0.5mm}{-0.5mm}{0mm}{black!10},
  fontupper=\small
}

% ----------------------------------------------------------------------
\subsection{Refinement Trace}
\label{app:qualitative}

Beyond the aggregate numbers, Figure~\ref{fig:qual-trace} follows the agent on the misuse rule, the only best-effort rule, from its first proposal to the record that Table~\ref{tab:titan-refinement} summarizes.
The agent's rationales are quoted verbatim, configurations use the paper's notation, and rules use the names of Appendix~\ref{app:constitution}.

\begin{figure}[t]
\begin{qbox}{\textsc{\textbf{\small Verification-Agent Trace}} \normalfont\footnotesize(misuse rule)}
\renewcommand{\arraystretch}{1.3}
\begin{tabularx}{\linewidth}{@{}>{\raggedright\arraybackslash}p{1.35cm}%
  @{\hspace{7pt}{\color{qline}\vrule width 0.8pt}\hspace{9pt}}%
  >{\raggedright\arraybackslash}p{1.55cm}%
  >{\raggedright\arraybackslash}X@{}}
\qstage{Rule} & \qfield{Instruction} & \qsay{For misuse-prone, malicious, or abusive requests, watermark the response and prioritize robustness over quality.} \\[5pt]
\qstage{Round 1} & \qfield{Proposal} & \qchip{$(\gamma,\delta,h,f) = (0.5,\,3,\,4,\,\texttt{min})$} \\
 & \qfield{Rationale} & \qsay{Robustness is the rule's top objective, so a stronger bias and a four-token minimum aggregation are selected. The minimum f-scheme aggregation can preserve the selected minimum when some recent-token inputs change, but this robustness rationale is a hypothesis requiring attack validation.} \\
 & \qfield{Result} & \qgate{Detection}{0.90}{\qno}\qdot\qgate{Quality}{1.00}{\qyes}\qdot\qgate{Robustness}{0.00}{\qno} \\[5pt]
\qdivider{intermediate rounds omitted} \\[5pt]
\qstage{Final} & \qfield{Proposal} & \qchip{$(\gamma,\delta,h,f) = (0.5,\,\mathbf{6},\,4,\,\texttt{min})$}\enspace{\scriptsize\color{qmuted}$\delta$ raised; $\gamma$, $h$, $f$ retained} \\
 & \qfield{Rationale} & \qsay{This is the specified unevaluated next configuration and follows the rule's robustness-first priority. It retains the best measured min/prefix structure and increases direct bias by one step as a robustness hypothesis, while all quality, detection, FPR, and robustness gates remain mandatory.} \\
 & \qfield{Result} & \qgate{Detection}{1.00}{\qyes}\qdot\qgate{Quality}{1.00}{\qyes}\qdot\qgate{Robustness}{0.10}{\qno} \\[5pt]
\qstage{Export} & \qfield{Status} & \qbesteffort \\
 & \qfield{Agent review} & \qsay{Strongest robustness-first incumbent: it has the highest measured robustness rate, 0.10 (5/50), while detection and quality both passed and detector-FPR regimes passed. Its higher PPL is secondary because robustness has priority. No measured candidate satisfies the robustness gate.} \\
\end{tabularx}
\end{qbox}
\caption{\textbf{Configuration refinement for the misuse rule.}
Results are measured on the 10-prompt reference set at nominal
$\alpha = 0.1\%$ against the gates detection $\geq 0.95$, quality
$\geq 0.90$, and robustness $\geq 0.50$; \qyes/\qno\ mark whether each is
met. Robustness counts detections over $50$ attacked responses
(10 prompts $\times$ 5 attacks), taking the worse of DIAA and DIPPER.}
\label{fig:qual-trace}
\end{figure}

\textbf{The agent grounds its proposals in the implementation and reports the gate it cannot meet.}
Its first proposal selects a four-token \texttt{min} aggregation for robustness from the KGW implementation and states that this rationale is a hypothesis that requires attack validation.
After the reference set rejected it on detection and robustness, the agent kept the \texttt{min} aggregation and the long context and raised the bias.
The final configuration passes detection and quality but reaches only 10\% robustness, so the agent exports it as best-effort and states that no measured candidate meets the robustness gate.

\subsection{Priority Following}
\label{app:priority-following}
Where the trace follows one rule, the sensitivity study of Table~\ref{tab:constitution-sensitivity} shows how the agent's choices change with the stated priority.
All three variants keep $\gamma=0.5$ and differ mainly in their bias.
This suggests that the agent searches a narrow region of KGW's configuration space rather than its full flexibility, which may explain why adjacent priorities are not reliably separated.
Changed parameters alone do not show that the constitution controls the watermark, so Table~\ref{tab:constitution-sensitivity} also reports the intended shift in measured outcomes.

\subsection{Deployment Routing}
\label{app:routing}
% Deployment-routing examples (Appendix C, Deployment Routing); styles are
% defined in qualitative_examples.tex, which is input earlier.
%
Finally, the deployment monitor decides which of these configurations a request receives.
Figure~\ref{fig:qual-routing} shows three decisions by \texttt{gpt-5-nano} at low reasoning effort, the second monitor of Table~\ref{tab:router-comparison}, chosen to contrast editing and advice requests on the same subject.
\textbf{The requested operation, not the subject, sets the policy.}
The monitor sends the request to edit a medical disclaimer to the editing exemption and watermarks the request for medical advice.
For a grammar correction annotated with both the editing and the deterministic-answer rule, it identifies only the deterministic-answer rule and watermarks the answer, although first-applicable precedence exempts it.
Omitting an applicable rule in this way is the failure mode behind this monitor's lower compliance in Table~\ref{tab:router-comparison}.

\begin{figure}[t]
\begin{qbox}{\textsc{\textbf{\small Deployment-Monitor Decisions}} \normalfont\footnotesize(\texttt{gpt-5-nano}, low reasoning)}
\renewcommand{\arraystretch}{1.3}
\begin{tabularx}{\linewidth}{@{}>{\raggedright\arraybackslash}p{2.35cm}>{\raggedright\arraybackslash}X@{}}
\multicolumn{2}{@{}p{\linewidth}@{}}{\qstage{Example 1}\enspace{\small Editing a medical text}\hfill\qcorrect} \\
\qfield{Prompt} & ``Rewrite the following medical disclaimer to improve readability: `This medication should only be taken as prescribed by a qualified healthcare professional.'\,'' \\
\qfield{Expected rule} & \qrule{Editing} $\rightarrow$ \qnowm \\
\qfield{Monitor decision} & \qrule{Editing} $\rightarrow$ \qnowm \\
\qfield{Rationale} & \qsay{The request is a low-risk proofreading/editing task; policy specifies no watermarking for low-risk edits.} \\[2pt]
\qsep \\
\multicolumn{2}{@{}p{\linewidth}@{}}{\qstage{Example 2}\enspace{\small Requesting medical advice}\hfill\qcorrect} \\
\qfield{Prompt} & ``I have a persistent cough and shortness of breath; what should I do?'' \\
\qfield{Expected rule} & \qrule{Advice} $\rightarrow$ \qwm \\
\qfield{Monitor decision} & \qrule{Advice} $\rightarrow$ \qwm \\
\qfield{Rationale} & \qsay{Medical advice with high-stakes implications; choose watermarking and prioritize quality/low distortion per policy.} \\[2pt]
\qsep \\
\multicolumn{2}{@{}p{\linewidth}@{}}{\qstage{Example 3}\enspace{\small Overlapping rules}\hfill\qwrong} \\
\qfield{Prompt} & ``Can you correct the grammar in this sentence: `She dont like the movie'?'' \\
\qfield{Annotated rules} & \qrule{Editing}\enspace\qrule{Deterministic} \\
\qfield{Expected rule} & \qrule{Editing} $\rightarrow$ \qnowm\enspace{\scriptsize\color{qmuted}first-applicable precedence} \\
\qfield{Monitor decision} & \qrule{Deterministic} $\rightarrow$ \qwm\enspace{\scriptsize\color{qmuted}editing rule not identified} \\
\qfield{Rationale} & \qsay{Deterministic grammar correction task; watermark applied per KGW with priority on quality and detectability.} \\
\end{tabularx}
\end{qbox}
\caption{\textbf{The same subject receives different policies when the requested operation differs.}
Examples 1 and 2 share a medical subject but differ in the requested
operation; Example 3 shows an omitted applicable rule. Prompts and
rationales are verbatim; rule identifiers are mapped to the names used in
Appendix~\ref{app:constitution}.}
\label{fig:qual-routing}
\end{figure}

\textbf{Exemption evasion.}
A misuse-prone user could frame a request as editing to obtain the editing exemption and an unwatermarked response.
We probe this with the \texttt{gpt-5.4-nano} monitor under the evaluated constitution, where first-applicable precedence places misuse before editing, so such requests should go to the misuse rule and receive a watermark.
The probes comprise 8 accepted corpus prompts that combine misuse and editing and 42 probes generated with an LLM from the same template, which skipped the critic but which the authors checked by hand.
The controls are 50 distinct corpus prompts that only request editing.
Table~\ref{tab:evasion} shows that the monitor routes every probe to the misuse rule and exempts 49 of the 50 controls, and its one control error watermarks an editing request, the conservative direction.
Because generated probes may carry more explicit misuse cues than real evasion attempts, we report the subsets separately and treat this as a pilot, and adaptive evasion remains future work.

\begin{table}[ht]
\centering
\small
\caption{\textbf{Exemption-evasion pilot} (\texttt{gpt-5.4-nano}, high reasoning).
Under first-applicable precedence, misuse precedes editing, so misuse requests framed as editing should be routed to the misuse rule and watermarked.
Brackets give two-sided 95\% Clopper--Pearson intervals. No routing failures occurred.}
\label{tab:evasion}
\vspace{2pt}
\setlength{\tabcolsep}{7pt}
\renewcommand{\arraystretch}{1.18}
\begin{tabular}{@{} l l c r @{}}
\toprule
\textbf{Probe set} & \textbf{Expected route} & $n$ & \textbf{Correct route (\%)}\,$\uparrow$ \\
\midrule
Misuse framed as editing: corpus prompts   & Misuse $\rightarrow$ WM & 8  & $100.0\;_{[63.1,\,100.0]}$ \\
Misuse framed as editing: generated probes & Misuse $\rightarrow$ WM & 42 & $100.0\;_{[91.6,\,100.0]}$ \\
\textbf{Misuse framed as editing: all}     & Misuse $\rightarrow$ WM & 50 & $\mathbf{100.0}\;_{[92.9,\,100.0]}$ \\
\addlinespace[2pt]
Editing-only controls (corpus)             & Editing $\rightarrow$ NoWM & 50 & $98.0\;_{[89.4,\,99.9]}$ \\
\bottomrule
\end{tabular}
\end{table}

\begin{table}[ht]
\centering
\small
\caption{\textbf{Routing on overlapping and single-rule test prompts.}
Overlapping prompts are annotated with more than one applicable rule.
\texttt{gpt-5-nano} results come from the end-to-end run in Table~\ref{tab:router-comparison}; \texttt{gpt-5.4-nano} results come from a separate routing-only rerun with the same instruction, frozen policy, and enforcer.
Decision accuracy is the resulting WM/NoWM decision over all prompts, distinct from the editing-only compliance in Table~\ref{tab:router-comparison}.
Routing failures count as incorrect; brackets give two-sided 95\% Clopper--Pearson intervals for rule assignment.}
\label{tab:overlap-routing}
\vspace{2pt}
\setlength{\tabcolsep}{5pt}
\renewcommand{\arraystretch}{1.18}
\begin{tabular}{@{} l l r c c r @{}}
\toprule
\textbf{Monitor} & \textbf{Prompts} & $n$ & \textbf{Correct rule (\%)}\,$\uparrow$ & \textbf{Correct decision (\%)}\,$\uparrow$ & \textbf{Failures}\,$\downarrow$ \\
\midrule
\texttt{gpt-5-nano} (low)    & Overlapping & 97  & $72.2\;_{[62.1,\,80.8]}$ & 93.8 & 4 \\
                             & Single-rule & 403 & $70.2\;_{[65.5,\,74.6]}$ & 88.6 & 32 \\
                             & All         & 500 & $70.6\;_{[66.4,\,74.6]}$ & 89.6 & 36 \\
\addlinespace[2pt]
\texttt{gpt-5.4-nano} (high) & Overlapping & 97  & $95.9\;_{[89.8,\,98.9]}$ & 97.9 & 0 \\
                             & Single-rule & 403 & $97.3\;_{[95.2,\,98.6]}$ & 99.0 & 0 \\
                             & All         & 500 & $97.0\;_{[95.1,\,98.3]}$ & 98.8 & 0 \\
\bottomrule
\end{tabular}
\end{table}

\textbf{Overlapping rules.}
Table~\ref{tab:overlap-routing} separates the 97 test prompts annotated with more than one applicable rule from the single-rule prompts.
For both monitors, accuracy on overlapping prompts is close to that on single-rule prompts, so the weaker monitor's errors do not concentrate on overlaps.
In a separate routing-only rerun, \texttt{gpt-5.4-nano} assigns 97.0\% of rules correctly, against 100.0\% in the end-to-end run of Table~\ref{tab:router-comparison}, which shows run-to-run variation in LLM routing that providers should monitor in deployment.

% ======================================================================
\section{Framework Extensions}
\label{app:extensions}
This appendix describes three extensions that the framework supports but our experiments do not evaluate, namely (i) a randomized-key forgery defense, (ii) specialist training of both agents and (iii) detection without the original prompt.

\subsection{Conditional Forgery Resistance}
\label{app:forgery}

We describe an optional composition with the randomized-key defense of
\citet{aremu2025randomizedkeys}, not a new security theorem or an evaluated
component of Constitutional-KGW. The argument applies to compatible
statistical watermarks under the assumptions below.

\textbf{Profile-conditioned construction.}
Fix a watermarked profile $c$, its policy version, and the selected algorithm
and configuration $(a_c,\psi_c)$. The provider holds $K\geq2$ secret,
compatible pairs of generation and detection keys
$\{(\mathsf{gk}_{c,j},\mathsf{dtk}_{c,j})\}_{j=1}^{K}$.
For every request assigned to $c$, independently sample
$J\sim\operatorname{Uniform}(\{1,\ldots,K\})$ and generate
\[
    \mathsf{Watermark}_{\mathsf{gk}_{c,J}}^{\mathcal{M}}
    (\pi;a_c,\psi_c)\to x.
\]
Only one key is embedded in each response. Detection tests all $K$ keys:
\begin{equation}
\begin{aligned}
    Z_{c,j}(x)
    &=\mathbf{1}\!\left[
      s_{a_c,\psi_c,\mathsf{dtk}_{c,j}}(x)\geq\eta_{c,j}\right],\\
    S_c(x)&=\sum_{j=1}^{K}Z_{c,j}(x),
    \qquad
    \mathsf{Acc}_c(x)=\mathbf{1}[S_c(x)=1].
\end{aligned}
    \label{eq:multikey-acceptance}
\end{equation}
Zero or multiple detections are rejected; multiple detections alone are not
proof of forgery. Exempt profiles remain unwatermarked and receive no
watermark-based provenance claim.

\textbf{Threat model and assumptions.}
The attacker knows the algorithm, configuration, and defense, may adaptively
collect $N$ provider responses, but knows neither the secret keys nor their
assignments and has no detection-oracle access. Target content
$\mathcal{H}_c$ lies outside legitimate provider output; the cited work uses
harmful content that the provider refuses to generate. Replaying a legitimate
oracle response is not a successful forgery. The following conditions delimit
the claim.

\emph{Fixed detection context.}
The profile and policy version are known through trusted metadata, or fixed
for the security analysis. Their algorithm, detector parameters, key set,
and thresholds are bound before testing the candidate text.

\emph{Key blindness.}
The attacker cannot reliably distinguish or partition collected responses
by their generating key, including through side information. Withholding key
labels alone does not establish this condition.

\emph{Adversarial-output symmetry and independence.}
For each attacker covered by the claim, its output $X^\star$ induces mutually
independent indicators
$Z_{c,j}(X^\star)\sim\operatorname{Bernoulli}(p)$ with the same marginal
$p\in[0,1]$ across keys. Here $p$ is attacker-induced, not the null FPR.
Key blindness does not imply this joint distribution, and independence on
benign null text does not establish independence on adversarial outputs.

\textbf{Conditional bound.}
Under these assumptions, the result of \citet{aremu2025randomizedkeys} yields
\begin{equation}
\begin{aligned}
    \operatorname{Adv}^{\mathrm{forge}}_c(\mathcal{B},N)
    &\leq\Pr[S_c(X^\star)=1]\\
    &=Kp(1-p)^{K-1}
    \leq\left(1-\frac{1}{K}\right)^{K-1}.
\end{aligned}
    \label{eq:conditional-forgery-bound}
\end{equation}
\emph{Proof.}
The independent, equal-marginal indicators give
$S_c(X^\star)\sim\operatorname{Binomial}(K,p)$.
For $0<p<1$, the derivative of $Kp(1-p)^{K-1}$ is
$K(1-p)^{K-2}(1-Kp)$, so its maximum occurs at $p=1/K$;
both endpoints give zero. Requiring $X^\star\in\mathcal{H}_c$ can only
decrease acceptance probability. $\square$

The bound does not depend on the number of collected samples $N$, provided
the assumptions continue to hold. It is per forgery attempt, not a bound on
success over unlimited attempts. Its value is $1/2$ for $K=2$ and tends to $1/e$ as $K$ grows. It is therefore not negligible, does not scale as $1/K$ and does not prove an improvement over every single-key attack success rate. It instead
excludes near-certain success within the stated attack model. The experiments
in the cited work are evidence for that work's evaluated settings, not for
our constitutional policies.

\textbf{False-positive calibration.}
Let $\alpha_{\mathrm{fw}}$ be the provider's overall null error budget and
$\alpha_{\mathrm{key}}$ a per-key level. For independent null indicators with
this common level, the \v{S}id\'ak choice used by the defense is
\begin{equation}
    \alpha_{\mathrm{key}}
    =1-(1-\alpha_{\mathrm{fw}})^{1/K},
    \qquad
    \Pr_0[S_c\geq1]=\alpha_{\mathrm{fw}}.
    \label{eq:multikey-null-calibration}
\end{equation}
The exactly-one acceptance FPR is at most this family-wise rate. Without
null independence, valid per-key levels at most
$\alpha_{\mathrm{fw}}/K$ instead give
$\Pr_0[\mathsf{Acc}_c=1]\leq\Pr_0[S_c\geq1]\leq\alpha_{\mathrm{fw}}$
by the union bound. These statements assume valid null levels; estimated
thresholds still require independent auditing and uncertainty reporting.
Neither calibration establishes the adversarial-output assumptions above.
Testing multiple profiles or choosing one from the candidate text requires calibrating the complete selection-and-detection procedure, because the fixed-profile bound does not transfer to an adaptive selector.

\textbf{Constitutional scope and limitations.}
A provider may request this defense for a forgery-sensitive rule, with its
assumptions recorded alongside the policy. Randomization must occur within
that profile; predictable selection between rules does not supply key
blindness. Mixing algorithms or configurations, including different
algorithms with the same key, does not automatically preserve equal attack
marginals or independence and requires a separate argument. Likewise,
distortion-freeness or public undetectability alone does not establish the
full set of assumptions. Although each response embeds only one watermark,
stricter thresholds and rejection of multiple detections can reduce genuine
acceptance, and testing $K$ keys increases detection cost.

Our experiments keep one key pair fixed and do not evaluate this wrapper,
its assumptions, or forgery attacks. We therefore record this extension as
\emph{conditional theoretical support}, not an empirically passed gate or a
security certificate for the deployed configurations.

\subsection{Specialist Training Objectives}
\label{app:training-objectives}
Hypothesis H6 of Appendix~\ref{app:hypotheses} posits that training both
agent roles for their tasks would improve performance. We give objectives for
this training; our experiments instantiate both roles with pretrained models
and do not optimize these objectives.
Verifier supervision spans algorithm implementations, configurations, proof
statements, checked derivations, and empirical failure cases across watermark
families. An example couples $v=(\Constitution,S,c,H_c)$ with a target $z^*$ that contains a legal algorithm and configuration, source-supported claims, explicit assumptions and predicted trade-offs. Here $S$ is the implementation bundle
and $H_c$ is the profile's measurement history. The supervised objective is
\begin{equation}
  \mathcal{L}_{\mathrm{ver}}(\theta)
  =-\mathbb{E}_{(v,z^*)\sim\mathcal{D}_{\mathrm{ver}}}
    \log A_\theta(z^*\mid v).
  \label{eq:verifier-training}
\end{equation}
Targets require independent review; unchecked model explanations cannot serve
as proof certificates. Subsequent preference learning can rank legal proposals
using measured constitutional attainment rather than explanation plausibility.

Monitor examples annotate every applicable rule, the resolved action, and
objective priorities. Let $y_r$ indicate whether rule $r$ applies, $w$ be the
correct watermark decision, and $p_{\varphi,r}$ be the predicted applicability
probability given $(\Constitution,\pi)$. One training objective is
\begin{equation}
  \mathcal{L}_{\mathrm{mon}}(\varphi)
  =\mathbb{E}_{\mathcal{D}_{\mathrm{mon}}}
    \left[\sum_{r\in\mathcal{S}}\operatorname{BCE}(y_r,p_{\varphi,r})
    -\log R_\varphi(w\mid \pi,\Constitution)\right].
  \label{eq:monitor-training}
\end{equation}
The training target must preserve multiple applicable rules, not replace them
with a single scenario label. Provider precedence and objective aggregation
are enforced outside the learned model. Confidence calibration and abstention
selection require held-out development data; classifier training alone does
not establish calibrated probabilities or reliable response-only recovery.

\subsection{Prompt-Free Detector Selection}
\label{sec:method-recovery}

Our evaluation uses known-configuration detection, which scores each generated response and its attacked variants with the detector bound to the configuration used for generation. This assumes a trusted generation record,
or authenticated metadata, identifying the policy version and detector; it is
not attribution from candidate text alone. We outline how the framework
extends to that setting.

\textbf{Response-based routing.}
Without the original prompt, the deployment monitor can infer a rule from the
visible response $x$, under a response-specific instruction, and retrieve that
rule's frozen detector. The monitor receives only $x$, the natural-language
rules, and the frozen policy; it sees neither the prompt nor private
reasoning, generation metadata, or detector scores. Its target is the rule
actually applied during generation, which need not be identifiable from $x$.
For KGW, the generation bias $\delta$ does not affect the score function, so
it suffices to recover a detector-equivalent configuration, which binds
the same $\mathsf{dtk}$, tokenizer and vocabulary size, $\gamma$, context
construction, and context length.
Let $j(x)$ index the selected detector, whose record binds
$(a_{j(x)},\psi_{j(x)},\mathsf{dtk}_{j(x)})$, and assign score $-\infty$ to
abstention or a \textsc{NoWM} decision. The complete procedure has score and
binary decision
\begin{equation}
  \begin{aligned}
    S_{\mathrm{mon}}(x)
      &=s_{a_{j(x)},\psi_{j(x)},\mathsf{dtk}_{j(x)}}(x),\\
    d_{\mathrm{mon}}^\alpha(x)
      &=\mathbf{1}[S_{\mathrm{mon}}(x)>\eta_{\mathrm{mon},\alpha}].
  \end{aligned}
  \label{eq:monitor-detection}
\end{equation}
A natural comparator instead scores the maximum over the unique deployed
detectors.

\textbf{Procedure-level calibration.}
Calibrating each detector individually does not calibrate text-dependent
selection, and a union of individually calibrated detections is not
calibrated at the same $\alpha$. Each complete procedure, including the
monitor, its instruction, and the policy, must therefore be frozen and then
calibrated on its own null scores. For a fixed procedure with score $S(x)$
and sorted null calibration scores $s_{(1)}\leq\cdots\leq s_{(n)}$, the
threshold at nominal rate $\alpha$ is
\begin{equation}
  k_\alpha=\left\lceil(n+1)(1-\alpha)\right\rceil,\qquad
  \eta_\alpha=
  \begin{cases}s_{(k_\alpha)},&k_\alpha\leq n,\\
  +\infty,&k_\alpha>n,
  \end{cases}
  \qquad d^\alpha(x)=\mathbf{1}[S(x)>\eta_\alpha].
  \label{eq:procedure-calibration}
\end{equation}
This order-statistic rule provides marginal null control under
exchangeability~\citep{angelopoulos2021conformal}, but not conditional FPR for
every realized calibration set or shifted domain. An untouched null audit
would then report the observed FPR with exact two-sided intervals and a
one-sided upper bound~\citep{clopper1934confidence}, together with selection
coverage, so that abstention cannot masquerade as a low-FPR detector. Rule
recovery, detector equivalence, and detection power are distinct outcomes,
particularly after paraphrasing; we leave their evaluation to future work.

% ======================================================================
\section{Extended Discussion and Limitations}
\label{app:extended-discussion}

\subsection{Hypotheses for Improvement}
\label{app:hypotheses}
Several factors held fixed in our experiments may affect performance.

\textbf{(H1) Unambiguous rules.}
Phrases such as ``minimize unnecessary distortion'' or ``balance these objectives equally'' leave much to the agent's interpretation.
Operational definitions, such as explicit priority weights or quantitative targets per objective, may better separate neighboring priorities.

\textbf{(H2) Theoretical and empirical context.}
The agent reasons from the watermark implementation and its own reference measurements.
Access to theoretical analyses and published trade-off curves may yield stronger first proposals and fewer wasted rounds.

\textbf{(H3) Rule-specific starting points.}
In the refinement trace of Appendix~\ref{app:qualitative}, the first proposal failed every robustness check before feedback corrected it.
Starting points supplied by the provider or validated earlier could reach passing or stronger best-effort policies within the same budget.

\textbf{(H4) Provider-specified legal bounds.}
Parameter ranges for each rule, such as an upper bound on the bias for quality-prioritized rules and a lower bound for robustness-prioritized rules, would narrow the search and exclude proposals that are legal for the watermark but implausible for the rule.

\textbf{(H5) Watermark diversity.}
Letting the agent choose among watermark families, rather than only among KGW configurations, may raise robustness beyond what the tested KGW configurations reach.

\textbf{(H6) Specialist training.}
Both agents are pretrained models used without weight updates, and we hypothesize that training them for their roles would improve performance.
A verification agent supervised on watermark implementations, checked derivations and failure cases may propose stronger first configurations and need fewer rounds.
A monitor trained on labels with several rules may route more reliably, and Table~\ref{tab:router-comparison} shows that routing errors lower detection, quality and compliance.
Appendix~\ref{app:training-objectives} gives training objectives for both roles.

\subsection{Limitations}
\label{app:limitations}
\textbf{Scope.} Our evaluation uses one watermark family (KGW), one serving model, one key pair, one search seed, and a single five-rule constitution. Generalization across these dimensions and the stability of search across seeds remain untested.

\textbf{Agents.} Both agents are pretrained models used without specialist training, so our results do not test hypothesis H6 or the objectives in Appendix~\ref{app:training-objectives}. Proposal budgets are matched with the random-search verifier, but compute cost is not, and we did not evaluate a deterministic heuristic or Bayesian optimization baseline.

\textbf{Finite-sample evidence.} Configurations are selected on $10$ reference prompts per rule, and the same null texts inform both selection and reported FPRs. Adaptive reuse can overfit these sets, so a policy may pass search and still fail at deployment, and reported intervals are descriptive rather than guarantees. Failing to find a passing configuration within the budget does not prove that the constitution is infeasible.

\textbf{Measurement.} Perplexity measures fluency, not task correctness or distribution preservation. The two paraphrasers do not exhaust possible attacks, and we do not audit whether attacked outputs preserve meaning. Prompts and rule labels are generated synthetically and require independent validation, particularly for overlapping or adversarial requests.

\textbf{Security.} We do not empirically evaluate forgery resistance. When all assumptions of \citet{aremu2025randomizedkeys} hold, randomly selecting among $K$ keys within each rule bounds a blind attacker's per-attempt forgery success, as Appendix~\ref{app:forgery} shows, and our framework can record this defense as part of a rule's policy. Rule-specific configurations under a single key do not by themselves satisfy these assumptions. Because users choose their prompts, an attacker can target a specific rule and collect samples from its configuration alone, and configurations sharing a key produce correlated rather than independent detection outcomes. Tuning KGW hyperparameters alone also does not establish distortion-freeness or undetectability.

\end{document}